\documentclass{llncs}

\usepackage{hyperref}
\usepackage{graphicx}
\usepackage{subfigure}

\newtheorem{finding}{Finding}

\begin{document}

\title{Computation of the Travelling Salesman Problem by a Shrinking Blob}
\titlerunning{Travelling Salesman Problem by a Shrinking Blob}

\author{Jeff Jones and Andrew Adamatzky}
\authorrunning{J. Jones and A. Adamatzky}

\institute{Centre for Unconventional Computing\\University of the West of England\\Coldharbour Lane\\Bristol, BS16 1QY, UK.\\
\email{jeff.jones@uwe.ac.uk, andrew.adamatzky@uwe.ac.uk}}

\maketitle

\begin{abstract}
The Travelling Salesman Problem (TSP) is a well known and challenging combinatorial optimisation problem. Its computational intractability has attracted a number of heuristic approaches to generate satisfactory, if not optimal, candidate solutions. Some methods take their inspiration from natural systems, extracting the salient features of such systems for use in classical computer algorithms. In this paper we demonstrate a simple unconventional computation method to approximate the Euclidean TSP using a virtual material approach. The morphological adaptation behaviour of the material emerges from the low-level interactions of a population of particles moving within a diffusive lattice. A `blob' of this material is placed over a set of data points projected into the lattice, representing TSP city locations, and the blob is reduced in size over time. As the blob shrinks it morphologically adapts to the configuration of the cities. The shrinkage process automatically stops when the blob no longer completely covers all cities. By manually tracing the perimeter of the blob a path between cities is elicited corresponding to a TSP tour. Over 6 runs on 20 randomly generated datasets of 20 cities this simple and unguided method found tours with a mean best tour length of 1.04, mean average tour length of 1.07 and mean worst tour length of 1.09 when expressed as a fraction of the minimal tour computed by an exact TSP solver. We examine the insertion mechanism by which the blob constructs a tour, note some properties and limitations of its performance, and discuss the relationship between the blob TSP and proximity graphs which group points on the plane. The method is notable for its simplicity and the spatially represented mechanical mode of its operation. We discuss similarities between this method and previously suggested models of human performance on the TSP and suggest possibilities for further improvement.
\end{abstract}

{\bf Keywords:} Travelling salesman problem, multi-agent, virtual material, unconventional computation, material computation

\section{Introduction}

The Travelling Salesman Problem (TSP) is a combinatorial optimisation problem well studied in computer science, operations research and mathematics. In the most famous variant of the problem a hypothetical salesman has to visit a number of cities, visiting each city only once, before ending the journey at the original starting city. The shortest path, or tour, of cities, amongst all possible tours is the solution to the problem. The problem is of particular interest since the number of candidate solutions increases greatly as $n$, the number of cities, increases. The number of possible tours can be stated as $(n-1)!/2$ which, for large numbers of $n$, renders assessment of every possible candidate tour computationally intractable. Besides being of theoretical interest, efficient solutions to the TSP have practical applications such as in vehicle routing, tool path length minimisation, and efficient warehouse storage and retrieval.

The intractable nature of the TSP has led to the development of a number of heuristic approaches which can produce very short --- but not guaranteed minimal --- tours. A number of heuristic approaches are inspired by mechanisms seen in natural and biological systems. These methods attempt to efficiently traverse the candidate search space whilst avoiding only locally minimal solutions and include neural network approaches (most famously in \cite{hopfield1986computing}), evolutionary algorithms \cite{larranaga1999genetic}, simulated annealing methods \cite{hasegawa2011verification}, the elastic network approaches prompted in \cite{durbin1987analogue}, ant colony optimisation \cite{dorigoa2000ant}, living \cite{aono_neurophys} and virtual \cite{Jones2011reconfig} slime mould based approaches, and bumblebee foraging \cite{lihoreau2010travel}. 

Human performance on the TSP has also been studied in both naive and tutored subjects (see, for example, \cite{macgregor2011human}). This is of particular interest because, unlike many nature inspired approaches, the human computation of TSP is by an individual and not based on population methods which evaluate a number of candidate solutions. Human performance on the TSP is also, for a limited number of cities at least, comparable in performance with heuristic approaches \cite{graham2000traveling}, \cite{dry2006human}. Although there are a number of competing theories as to how exactly humans approximate the TSP \cite{macgregor2000model},\cite{graham2000traveling}, \cite{pizlo2006traveling}, discovery of the methods employed may be useful as an insight into the mechanisms underlying complex perceptual and cognitive processes and potentially as an aid for the development of computational algorithms.

In this paper we adopt a material-based, minimum complexity approach. We show how a spatially represented non-classical, or unconventional, computational mechanism can be used to approximate the TSP. Taking inspiration from the non-neural, material-based computational behaviour of slime mould, we employ a sheet, or `blob' of virtual material which is placed over a spatial map of cities. By shrinking this blob over time, it conforms and adapts to the arrangement of cities and a tour of the TSP is formed. We give an overview of the inspiration for the method in Section \ref{sec:inspiration}. The shrinking blob method is described in Section \ref{sec:blobdesc}. Examples of the performance of the method compared to exact solutions generated by a TSP solver are given in Section \ref{sec:results}, along with an analysis of the underlying mechanism and factors affecting the performance of the approach. We conclude in Section \ref{sec:discussion} by summarising the approach and its contribution in terms of simplicity. We examine similarities between the underlying mechanism of the shrinking blob method and proposed models of TSP tour perception and construction in studies of human performance on the TSP. We suggest further research aimed at improving the method.

\section{Slime Mould Inspired Computation of the TSP}
\label{sec:inspiration}

The giant single-celled amoeboid organism, true slime mould \emph{Physarum polycephalum}, has recently been of interest as a candidate organism for the study of non-neural distributed computation. In the vegetative plasmodium stage of its complex life cycle the organism forages towards, engulfs and consumes micro-organisms growing on vegetative matter. When presented with a spatial configuration of nutrient sources the plasmodium forms a network of protoplasmic tubes connecting the nutrients. This is achieved without recourse to any specialised neural tissue. The organism dynamically adapts its morphology to form efficient paths (in terms of a trade-off between overall distance and resilience to random damage) between the food sources \cite{NakagakiT04MultFoodSrc},\cite{nakagaki2007intelligent},\cite{nakagaki2007effects}.

Research into the computational abilities of \emph{Physarum} was prompted by Nakagaki, Yamada and Toth, who reported the ability of the \emph{Physarum} plasmodium to solve a simple maze problem \cite{NakagakiT00MazeSolve}. It has since been demonstrated that the plasmodium successfully approximates spatial representations of various graph problems \cite{NakagakiT04MultFoodSrc,shirakawa2009planedivision,adamatzky_toussaint,jones2010influences}, combinatorial optimisation problems \cite{AonoM07PhysarumNeuroComp,aono2008spontaneous}, construction of logic gates and adding circuits \cite{TsudaS04PhysarumComp,jones2010towards,adamatzky_physarumgate}, and spatially represented logical machines \cite{adamatzky2007kum,adamatzky_jones_NC}. 

\begin{figure}[htbp]
 \centering
 \subfigure[0h]{\includegraphics[width=0.24\textwidth]{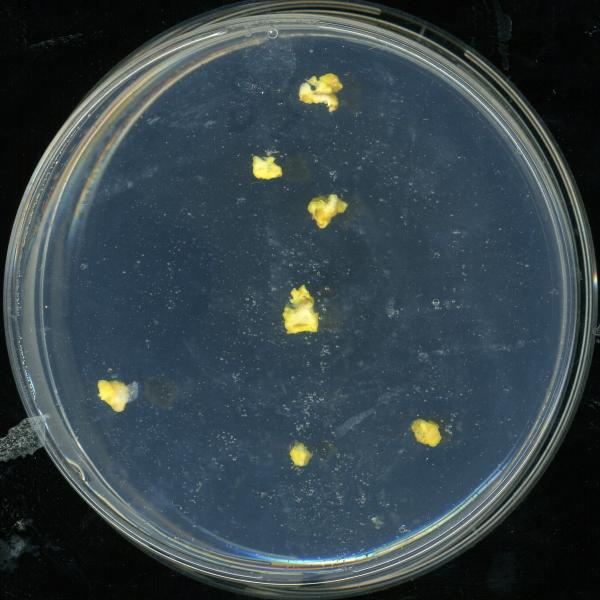}}
 \subfigure[12h]{\includegraphics[width=0.24\textwidth]{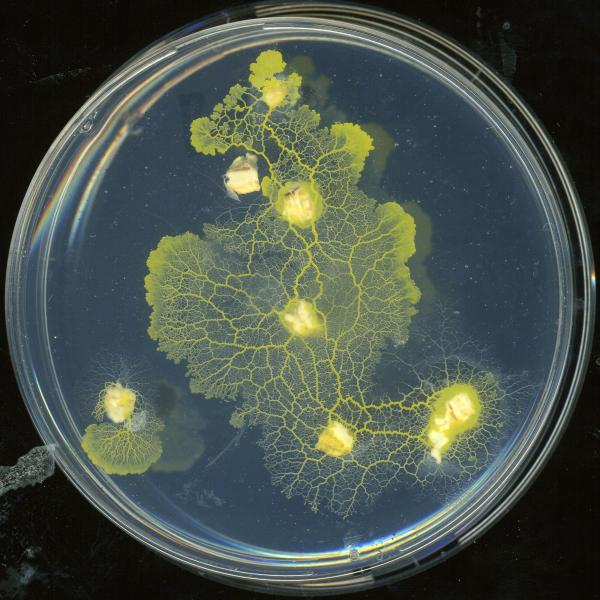}}
 \subfigure[24h]{\includegraphics[width=0.24\textwidth]{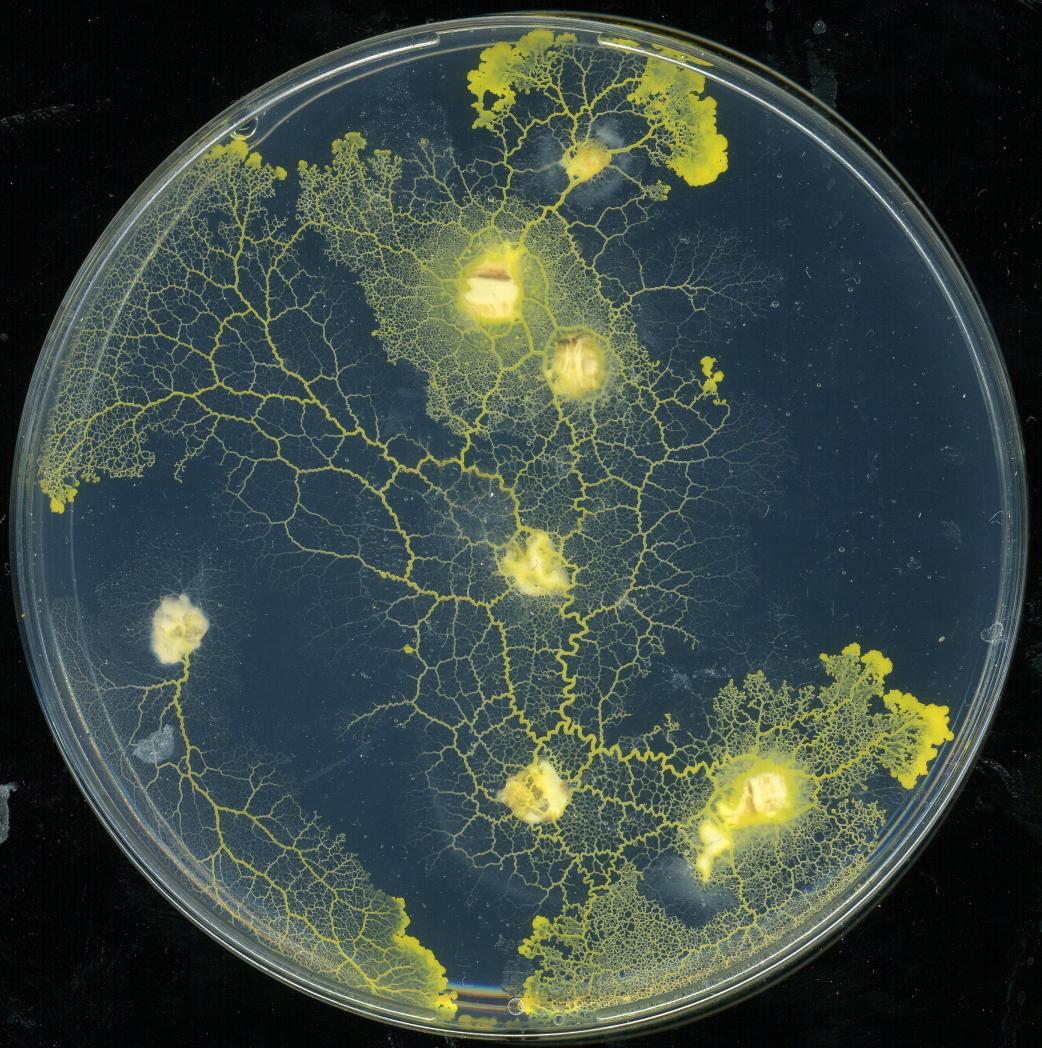}}
 \subfigure[36h]{\includegraphics[width=0.24\textwidth]{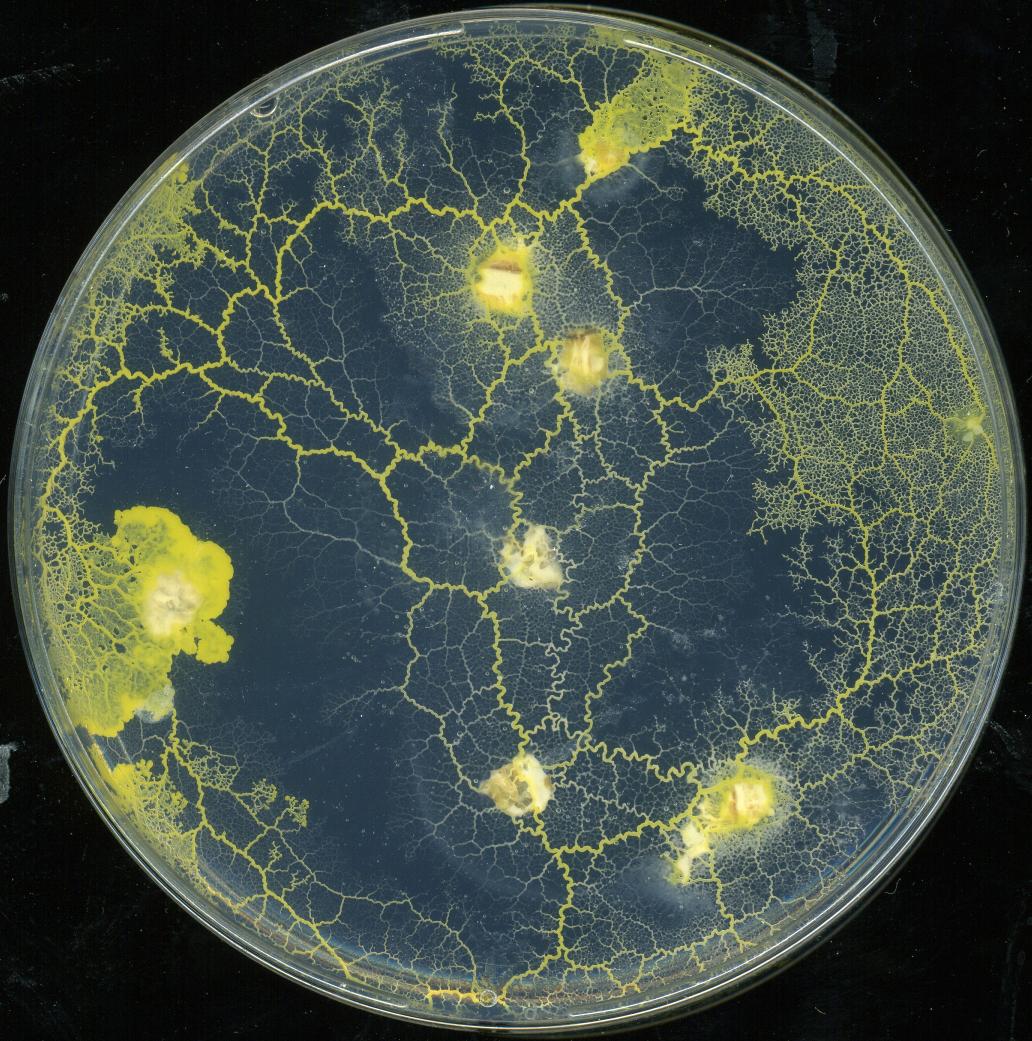}}   
 \subfigure[10h]{\includegraphics[width=0.32\textwidth]{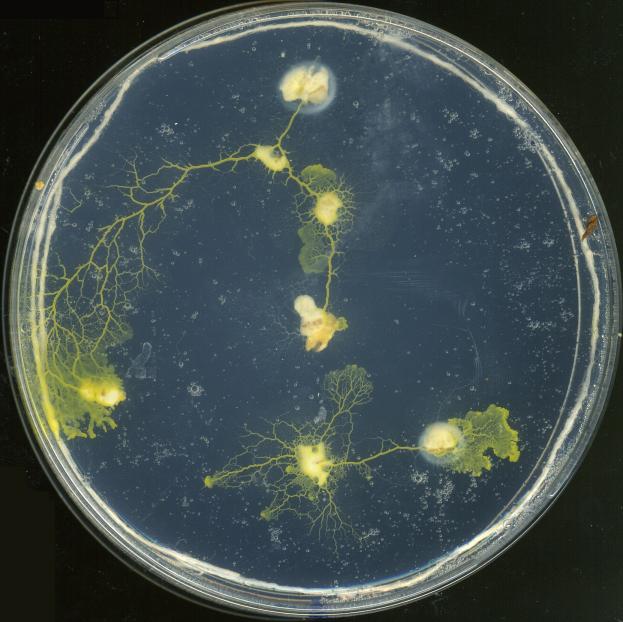}}
 \subfigure[24h]{\includegraphics[width=0.32\textwidth]{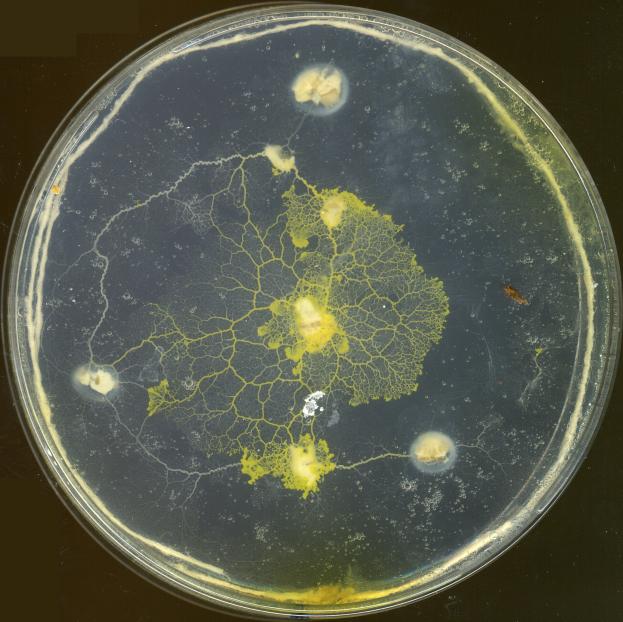}}
 \subfigure[36h]{\includegraphics[width=0.32\textwidth]{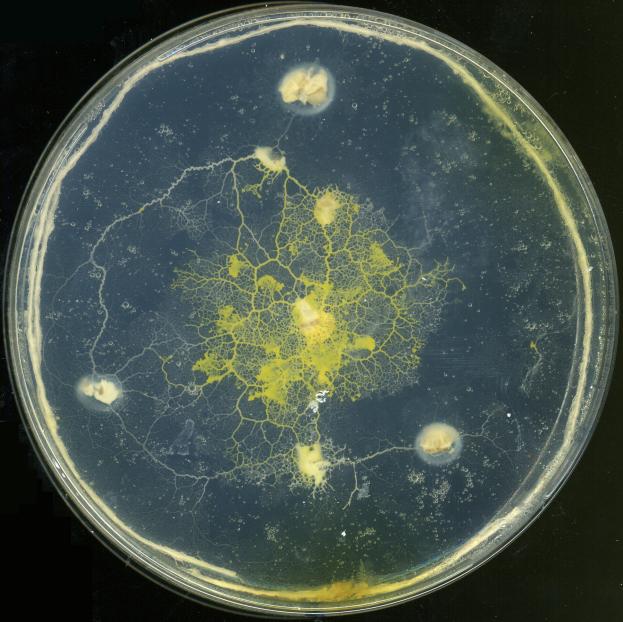}}  

\caption[]{Foraging plasmodium of \emph{Physarum} does not approximate the TSP in both unconstrained and constrained environments. (a) \emph{Physarum} plasmodia are inoculated at oat flakes on non-nutrient agar, (b) individual plasmodia extend from oat flakes and fuse, (c-d) the plasmodium continues to forage and the shape of the TSP is not represented. (e) foraging of plasmodium is constrained by placing a ring of saline soaked thread (1g NaCl / 100g water) at the periphery of the arena, (f) as the salt diffuses into the agar the shape of the plasmodium is confined, (g) the pattern of the plasmodium at 36h is confined but is disconnected from outer nodes (note empty tube remnants) and does not approximate the TSP.} 
\label{fig:physarum_attempt}
\end{figure}

Although \emph{Physarum} has been previously been used in the approximation of TSP \cite{aono_neurophys}, this was achieved by an indirect encoding of the problem representation to enable it to be presented to a confined plasmodium in a controlled environment. In the work by Aono et. al. it was shown that the morphology of the plasmodium confined in a stellate chamber could be dynamically controlled by light irradiation of its boundary. When coupled to an elegant feedback mechanism using an analysis method (to assess the presence of plasmodium at the extremities of the chamber), combined with Hopfield-Tank type neural network rules \cite{hopfield1986computing}, the plasmodium was used to generate candidate solutions to simple instances  of the TSP \cite{AonoM07PhysarumNeuroComp,aono2008spontaneous}.In its natural propagative state, however, \emph{Physarum} does not approximate area representations of a set of points, including the Convex Hull, Concave Hull \cite{adamatzky2011planarshape} and the TSP. This is because the material comprising the plasmodium spontaneously forms networks spanning the nutrient sources. Even when the plasmodium is arranged initially as a solid sheet of material, the sheet is soon transformed into a network structure by competitive flux of material within the sheet \cite{NakagakiT04MultFoodSrc}. It is physically impractical to force a freely foraging plasmodium to conform to a TSP network structure during its nutrient foraging, as shown in Fig. \ref{fig:physarum_attempt}. 

Nevertheless, the material computation embodied within \emph{Physarum} presents interesting possibilities towards generating novel spatially represented methods of unconventional computation. The motivation for this is threefold. Firstly, many natural systems exhibit properties which are not found in classical computing devices, such as being composed of simple and plentiful components, having redundant parts (i.e. not being dependent on highly complex units), and showing resilient, or fault tolerant, behaviour. Secondly, unconventional computation is often observed in systems which show emergent behaviour, i.e. behaviour which emerges from the interactions between simple component parts, and which --- critically --- cannot be described in terms of the lower level component interactions. Emergent behaviour is characterised by systems with many simple, local interactions and which display self-organisation --- the spontaneous appearance of complexity or order from low-level interactions. The study of these properties is useful not only from a computational perspective, but also from a biological viewpoint -- since much of the complexity in living systems appears to be built upon these principles. The third reason for interest in unconventional computation is because, for a number of applications at least, utilising the natural properties of physical systems is a much more efficient means of computation. Natural computation can take advantage of parallel propagation of information through a medium (for example in the chemical approximation of Voronoi diagrams (\cite{de2004formation}), or the parallel exploration of potential path choices in path problems using microfluidic gas discharge plasmas (\cite{reyes2002glow}). Stepney has suggested that this \emph{material computation}, in utilising the natural physical properties of materials (not necessarily living systems), may afford rich computational opportunities \cite{stepney2008neglected}.

An \emph{in-silico} attempt at reproducing the pattern formation and adaptation behaviour of \emph{Physarum} using multi-agent transport networks was introduced in \cite{jones2010emergence} and its approximation of TSP paths for very simple data layouts was noted. The general pattern formation abilities of the approach was characterised in \cite{jones_alife_2010}. Attempts at encouraging these multi-agent transport networks to conform to TSP-like requirements (degree of connectivity 2, no crossed paths) by dynamically adjusting the concentration of simulated nutrient attractants, using a feedback mechanism based on the current configuration of the network, were presented in \cite{Jones2011reconfig}. This resulted in extremely complex transitions of network dynamics and partial success in constructing TSP tours. In the approach outlined in this paper we attempt a simpler approach which utilises a larger aggregate mass of the same multi-agent collective which behaves as a morphologically adaptive cohesive `blob' of virtual material.

%
%

\section{A Material Approach to TSP by a Shrinking Blob}
\label{sec:blobdesc}

In the shrinking blob method we use a piece, or `blob' of a virtual plasmodium material to approximate the TSP. The material is composed of thousands of simple mobile multi-agent particles interacting together in a 2D diffusive lattice. Each particle senses the concentration of a generic `chemoattractant' diffusing within the lattice and each agent also deposits the same substance within the lattice upon successful forward movement. The multi-agent population collectively exhibits emergent properties of cohesion and shape minimisation as a results of the low-level particle interactions. The pattern formation and network adaptation properties of small populations of the material were discussed in \cite{jones_alife_2010} and were found to reproduce a wide range of Turing-type reaction-diffusion patterning. In this paper we use a relatively large population of particles which collectively behaves as a sheet of deformable virtual material. A full description of the virtual material method is given in the Appendix and an overview of the method follows.

\subsection{Shrinkage Process}

\begin{figure}[htbp]
 \centering
 \subfigure[t=60]{\includegraphics[width=0.29\textwidth]{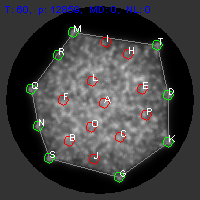}}
 \subfigure[t=60]{\includegraphics[width=0.29\textwidth]{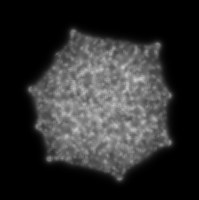}}
 \subfigure[t=520]{\includegraphics[width=0.29\textwidth]{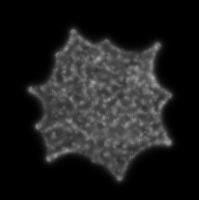}}
 \subfigure[t=3360]{\includegraphics[width=0.29\textwidth]{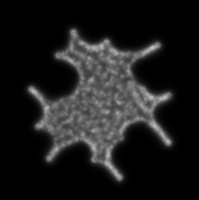}}
 \subfigure[t=6360]{\includegraphics[width=0.29\textwidth]{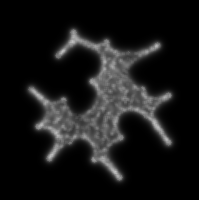}}
 \subfigure[t=7102]{\includegraphics[width=0.29\textwidth]{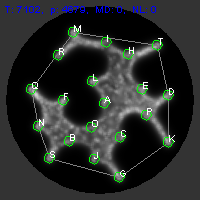}}
\caption[]{Visualisation of the shrinking blob method. (a) sheet of virtual material initialised within the confines of the convex hull (grey polygon) of a set of points. Node positions are indicated by circles. Outer partially uncovered nodes are light grey, inner nodes covered by the sheet are in dark grey, (b-e) sheet morphology during shrinkage at time 60, 520, 3360, 6360 respectively, (f) shrinkage is stopped automatically when all nodes are partially uncovered at time 7102.} 
\label{fig:shrink_method}
\end{figure}

We initialise a sheet of the virtual material around a set of data points corresponding to TSP city nodes (Fig. \ref{fig:shrink_method}a). Chemoattractant is projected into the diffusive lattice at node locations, however, projection is reduced at regions which are covered by the blob sheet. The initial shape of the sheet corresponds to the Convex Hull of the data points. We then shrink the material by systematically removing some of its constituent particle components. The city nodes act as attractants to the material, effectively `snagging' the material at the locations of uncovered nodes and affecting its subsequent morphological adaptation. As the material continues to shrink its innate minimising properties conform to the locations of the city nodes and the area occupied by the material is reduced, becoming a concave area covering the nodes (Fig. \ref{fig:shrink_method}b-e). The shrinkage is stopped when all of the nodes are partially uncovered by the sheet (Fig. \ref{fig:shrink_method}f). The reader is encouraged to view the supplementary video recordings of the shrinkage process at \url{http://uncomp.uwe.ac.uk/jeff/material_tsp.htm}. The adaptation of the blob to the data stimuli is not entirely smooth, the video recordings show that the blob sheet adapts to the changing stimuli as data nodes are temporarily uncovered and re-covered by the blob. When the shrinkage is halted the area of the sheet corresponds to the area enclosed by a tour of the Euclidean Travelling Salesman Problem. The exact tour formed by the blob can be elucidated by tracking along the perimeter of the blob, adding a city to the tour list when it is first encountered. The tour is complete when the start city is re-encountered. The approach is simple, making use of the innate adaptive emergent properties of the material. Despite being completely unguided and containing no heuristic optimisation strategies the approach yields efficient tours. The separate stages of the approach will now be described in detail.

\subsection{Halting the Computation}

It is important to halt the shrinkage of the blob at the right time. If the shrinking is stopped too early an incomplete tour will be formed (i.e. only a partial subset of the nodes will be included in the tour if not all of the nodes are uncovered). Unlike guided heuristic methods a set of candidate tours is not initially formed and subsequently modified. Only a single tour is formed and the shrinking blob approach is akin to the `instance machines' (as opposed to universal machines) proposed by Zauner and Conrad \cite{zauner1996parallel}. To automatically halt the computation we use a so-called `traffic light' system. At the start of the method the sheet covers the entire set of nodes. Only the outer nodes are partially covered by the blob. To measure whether a node is covered by the sheet we assess the number of particles in a $5x5$ window around each node. If the number of particles is $<15$ then the node is classified as uncovered and the node indicator is set to green. Otherwise the node is classified as covered and the node indicator is set to red. At each scheduler step the indicators of all nodes are checked. When all nodes are set to green, all nodes underneath the blob are partially uncovered and the shrinkage is stopped.

\subsection{Reading the Result of the Computation}
\label{read_results}

\begin{figure}[htbp!]
 \centering
 \includegraphics[width=0.6\textwidth]{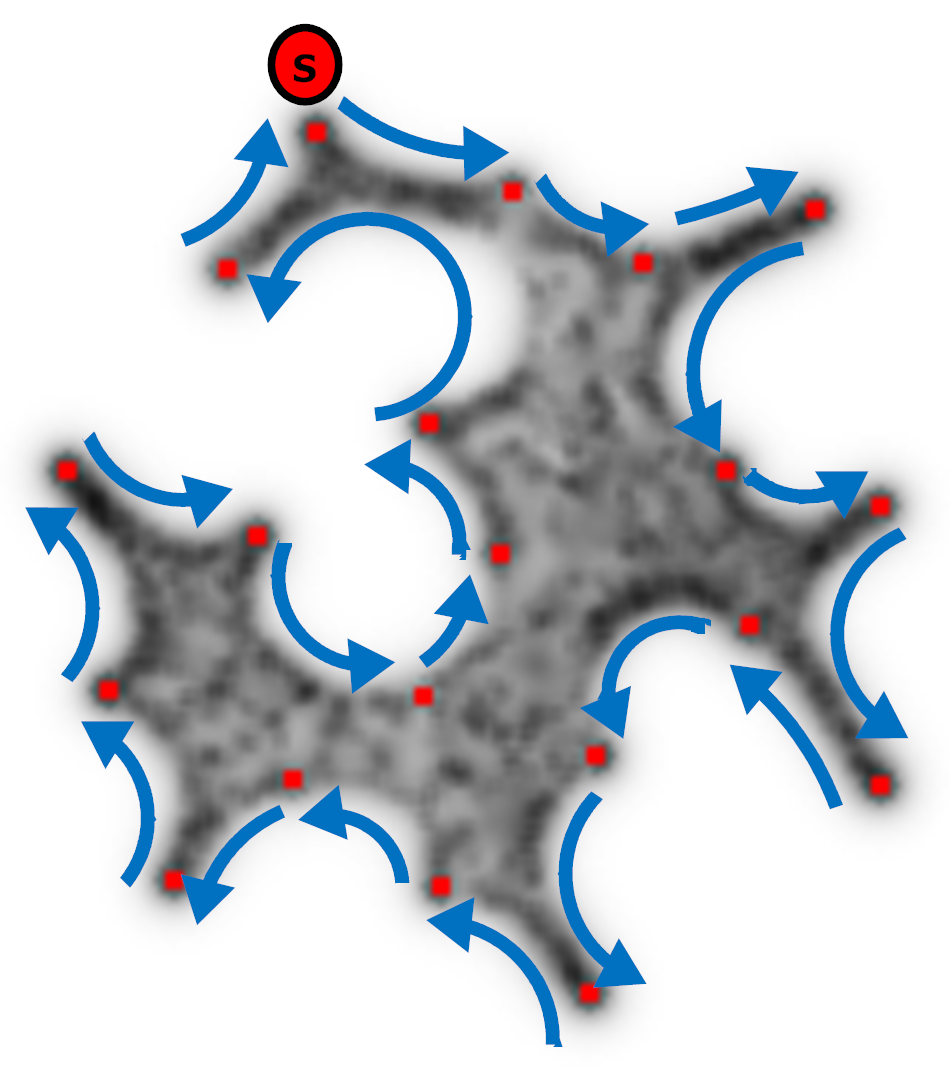}
\caption[]{Reading the TSP tour formed by the shrinking blob by perimeter tracking. (a) Tracking is initialised at the top most node. Perimeter of blob is traced in a clockwise direction. Each time a node is encountered for the first time it is added to the tour. The tour is completed when the start node is re-encountered.} 
\label{fig:trace_method}
\end{figure}

To trace the path of cities in the tour discovered by the blob a manual process is used. The collection of partially uncovered nodes and blob shape may be interpreted as an island shape with the nodes representing cities on the coastline of the island (Fig. \ref{fig:trace_method}). We begin by selecting the city at the top of the arena. If more than one city is at this $y$ location the left-most city at this $y$ location is selected. This city is the start city of the tour and is added to the tour list $\textbf{T}$. Moving in a clockwise direction we trace the perimeter of the blob (walking around the shore of the island \ldots). Each time we encounter a city, it is added to $\textbf{T}$. If a city is subsequently re-encountered (as in the case of narrow peninsula structures as described below) it is ignored. When the path reaches the starting city the tour is complete and the list in $\textbf{T}$ represents the tour of the TSP found by the shrinking blob.

\begin{figure}[htbp]
\centering
 \subfigure[]{\includegraphics[width=0.55\textwidth]{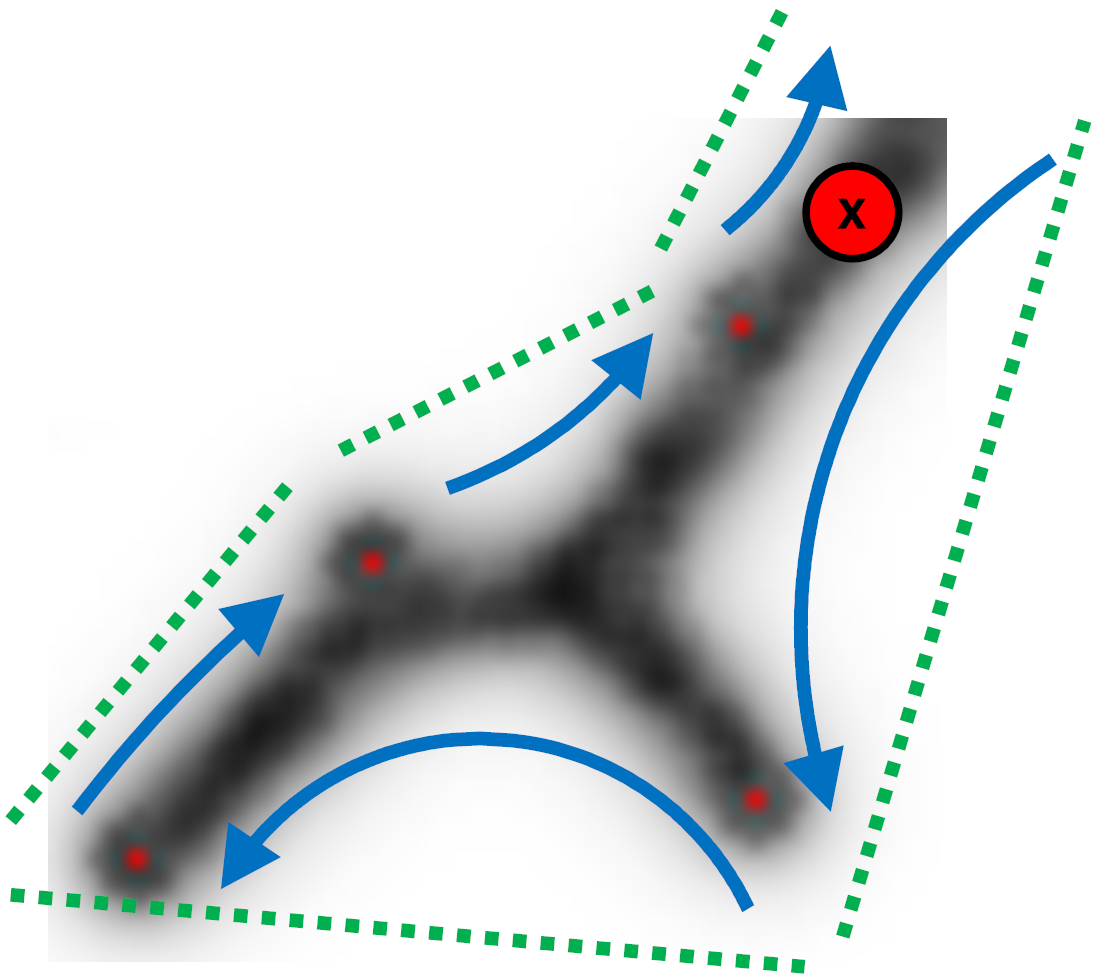}}
 \subfigure[]{\includegraphics[width=0.9\textwidth]{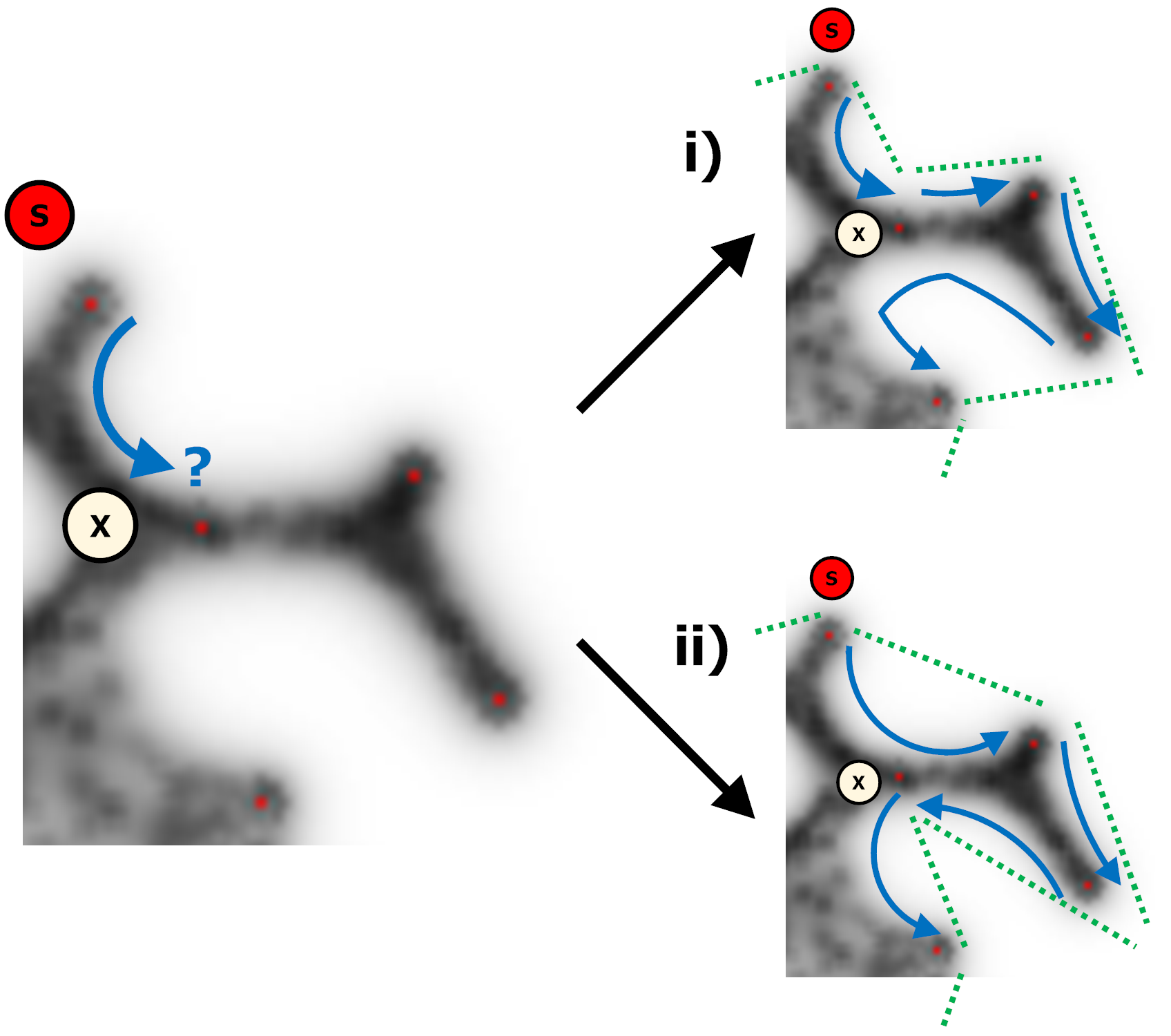}} 
\caption[]{ Special cases of when nodes are located on a narrow peninsula, close, or equidistant from either side of the `land'. (a) The node at `x' is close to the middle of a narrow portion of the blob. The slight convex bulge in the blob indicates that it is closest to the left side and the node is not added to the tour until it is encountered on the left side. (b) The node at `x' is directly in the middle of a narrow portion of the blob. Two potential tours are possible, shown in i) and ii) with their respective tours as dotted lines. If this case occurs, the node is added to the tour the first time it is encountered, as in i).}
\label{fig:peninsula_cases}
\end{figure}

Some special cases in the tracking process must be noted in the case where a city lies on a narrow `peninsula' of the blob as indicated in Fig. \ref{fig:peninsula_cases}. In Fig. \ref{fig:peninsula_cases}a the city nearest position $x$ in the path lies close to one side of a narrow peninsula. However the side at which the city is located can be deduced by a small convex bulge on the left side of the blob. In this case the city is not added until it is encountered on the left side of the peninsula. In the case of Fig. \ref{fig:peninsula_cases}b, however, the city at $x$ is located exactly in the middle of a peninsula and its closest side cannot be discerned. In this instance two interpretations are possible and the subsequent differences in possible tour paths are indicated by the dotted lines in Fig. \ref{fig:peninsula_cases}b, i) and ii). In interpretation i) the city is added to $\textbf{T}$ immediately and in ii) it is not added until it is encountered on its opposite side. If this situation occurs during the tracking process we add the city to $\textbf{T}$ when it is first encountered.

\section{Results}
\label{sec:results}

\begin{figure}[htbp]
 \centering
 \includegraphics[width=0.95\textwidth]{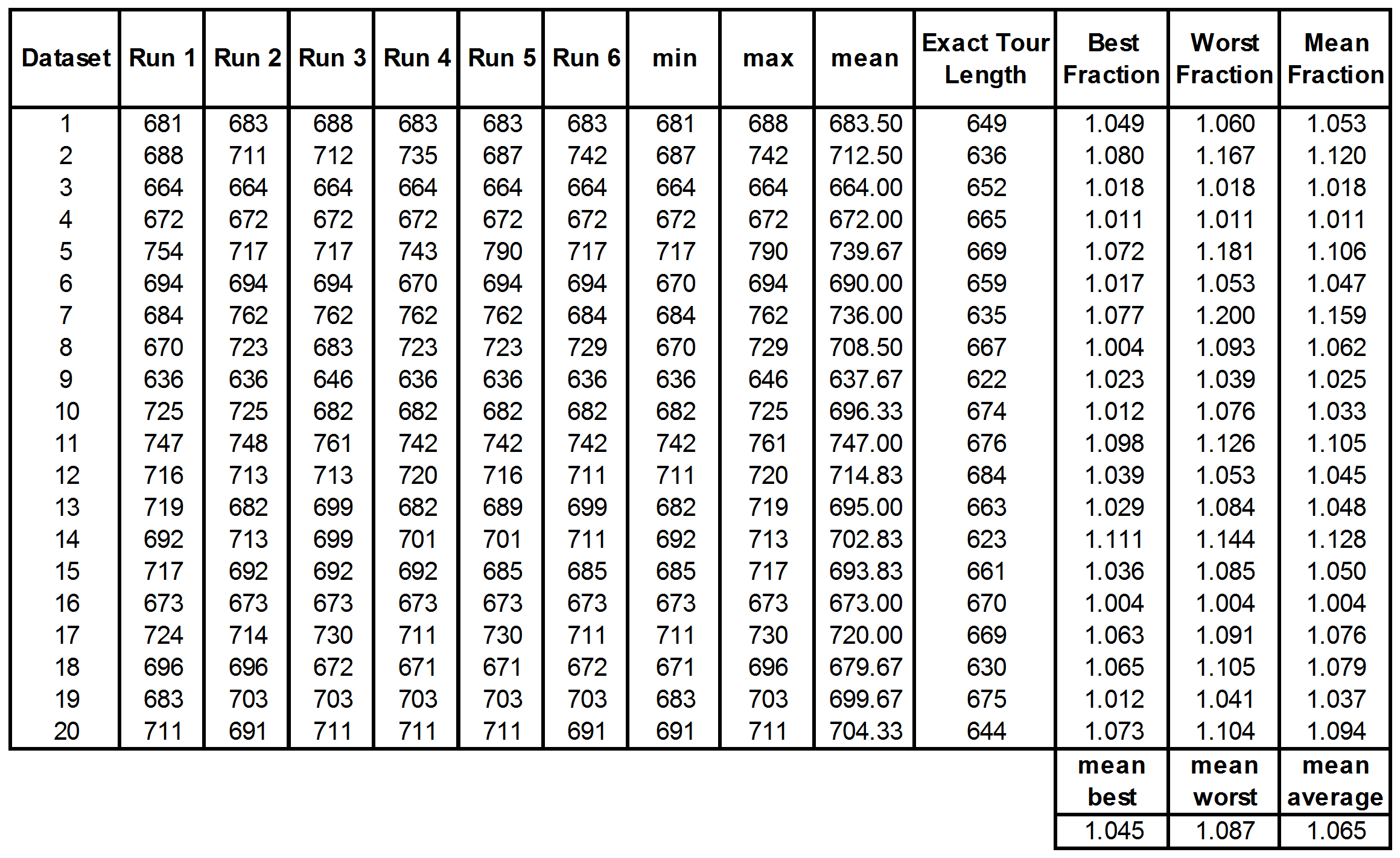}
\caption[]{Results of shrinking blob method over 6 runs on each of 20 randomly generated datasets of 20 points compared to exact results from the Concorde TSP solver.} 
\label{fig:results}
\end{figure}

We assessed the shrinking blob method by generating 20 datasets, each consisting of 20 randomly generated nodes within a circular arena in a $200x200$ lattice. To aid the manual tracking process we added the condition that points must have a separation distance of at least 25 pixels. For each run a population of particles was generated and initialised within the confines of the convex hull (algorithmically generated) of the point set. Any particles migrating out of the convex hull area were removed. As the shrinkage process started the cohesion of the blob emerged and, as shrinkage progressed, the blob adapted to the shape of the city nodes. Six experimental runs were performed on each dataset and the resulting blob shape was recorded and tracked by the manual tracking process to reveal the tour. The best, worst and mean performance over the 6 runs for each 20 datasets was recorded and these results were aggregated over the 20 datasets and shown in Fig. \ref{fig:results}. Results of the shrinking blob method are expressed as a fraction of the shortest exact tour found by the Concorde TSP solver \cite{applegate2006concorde}.

\begin{figure}[htbp]
 \centering
 \subfigure[Blob 670 (1.004)]{\includegraphics[width=0.35\textwidth]{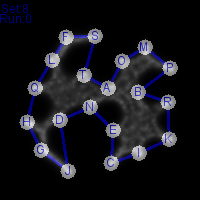}}
 \subfigure[Concorde 667]{\includegraphics[width=0.35\textwidth]{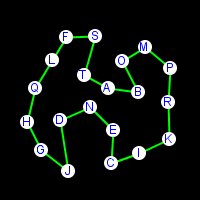}} 
 \subfigure[Blob 672 (1.011)]{\includegraphics[width=0.35\textwidth]{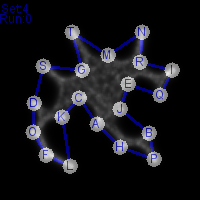}}
 \subfigure[Concorde 665]{\includegraphics[width=0.35\textwidth]{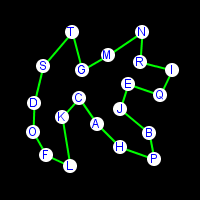}}
 \subfigure[Blob 673 (1.004)]{\includegraphics[width=0.35\textwidth]{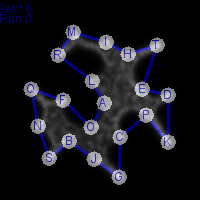}}
 \subfigure[Concorde 670]{\includegraphics[width=0.35\textwidth]{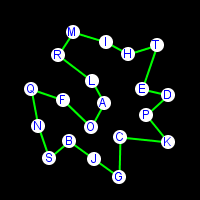}}
 \subfigure[Blob 683 (1.012)]{\includegraphics[width=0.35\textwidth]{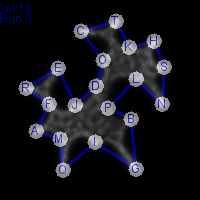}}
 \subfigure[Concorde 675]{\includegraphics[width=0.35\textwidth]{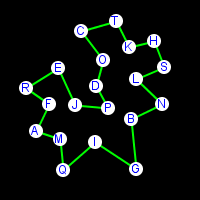}}   
\caption[]{Examples of good performance by the shrinking blob method. (a,c,e,g) Final blob shape with TSP tour overlaid, tour length and fraction of exact tour in parentheses, (b,d,f,h) Minimum exact tour found by the Concorde TSP solver.} 
\label{fig:results_good}
\end{figure}

\begin{figure}[htbp]
 \centering
 \subfigure[Blob 742 (1.167)]{\includegraphics[width=0.35\textwidth]{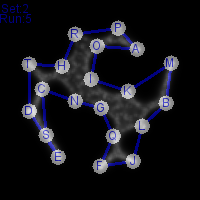}}
 \subfigure[Concorde 636]{\includegraphics[width=0.35\textwidth]{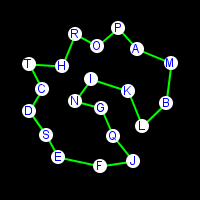}}
 \subfigure[Blob 762 (1.20)]{\includegraphics[width=0.35\textwidth]{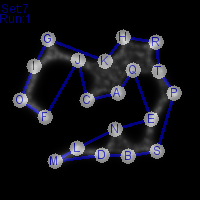}}
 \subfigure[Concorde 635]{\includegraphics[width=0.35\textwidth]{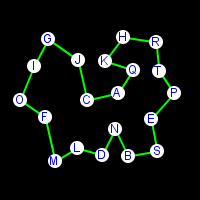}}
 \subfigure[Blob 761 (1.126)]{\includegraphics[width=0.35\textwidth]{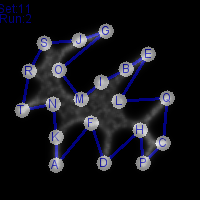}}
 \subfigure[Concorde 676]{\includegraphics[width=0.35\textwidth]{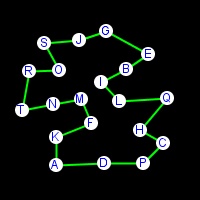}}
 \subfigure[Blob 713 (1.144)]{\includegraphics[width=0.35\textwidth]{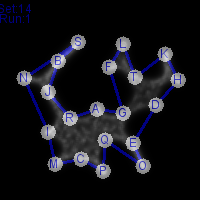}}
 \subfigure[Concorde 623]{\includegraphics[width=0.35\textwidth]{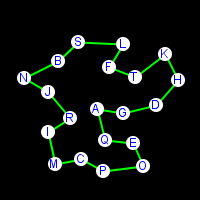}}   
\caption[]{Examples where shrinking blob does not approximate TSP well. (a,c,e,g) Final blob shape with TSP tour overlaid, tour length and fraction of exact tour in parentheses, (b,d,f,h) Minimum exact tour found by the Concorde TSP solver.} 
\label{fig:results_not_good}
\end{figure}

\subsection{Construction of Tour by Concave Insertion Process}

\begin{figure}[htbp]
 \centering
 \subfigure[]{\includegraphics[width=0.24\textwidth]{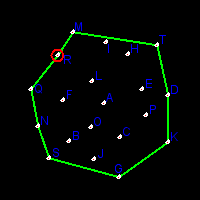}}
 \subfigure[]{\includegraphics[width=0.24\textwidth]{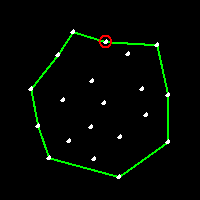}}
 \subfigure[]{\includegraphics[width=0.24\textwidth]{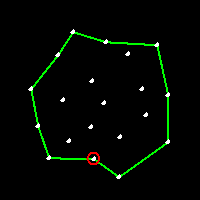}}
 \subfigure[]{\includegraphics[width=0.24\textwidth]{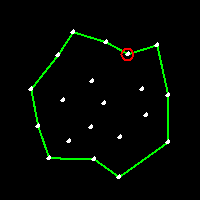}}
 \subfigure[]{\includegraphics[width=0.24\textwidth]{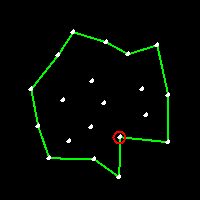}}
 \subfigure[]{\includegraphics[width=0.24\textwidth]{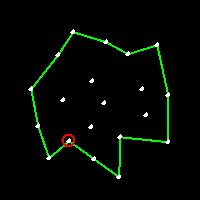}}
 \subfigure[]{\includegraphics[width=0.24\textwidth]{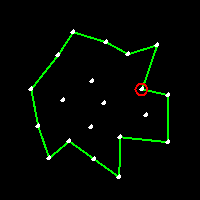}}
 \subfigure[]{\includegraphics[width=0.24\textwidth]{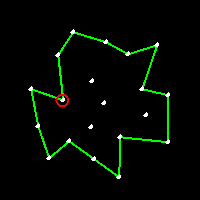}}
 \subfigure[]{\includegraphics[width=0.24\textwidth]{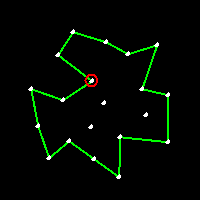}}     		        \subfigure[]{\includegraphics[width=0.24\textwidth]{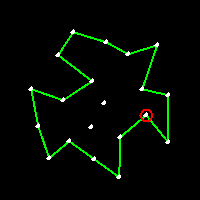}}
 \subfigure[]{\includegraphics[width=0.24\textwidth]{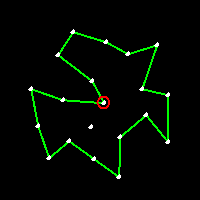}}
 \subfigure[]{\includegraphics[width=0.24\textwidth]{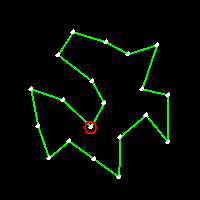}}            
 \caption[]{Construction of TSP tour by shrinking blob includes the transformation between Convex Hull and Concave Hull. (a) Initial Convex Hull of dataset 16 (shown by points connected by path) is deformed to a concave shape by the shrinking blob. The stepwise construction of the tour is indicated by adding a circled point as each new city is discovered, (b-k) As the blob continues to shrink new points are included (circled) further reducing the area of the Concave Hull, (l) when shrinkage stops the set of encompassed points is a tour of the TSP.} 
\label{fig:conv_conc_tsp}
\end{figure}

Although the final tour list is read off by tracking the perimeter of the shrunken blob, the construction of the tour actually occurs by an insertion process as the blob shrinks. The blob is initially patterned with the shape of the convex hull. This is only a partial tour, since only the peripheral nodes which are part of the Convex Hull are included. By recording the stages by which nodes are uncovered and added during the shrinkage process, the method of construction can be elucidated. Fig.~\ref{fig:conv_conc_tsp} shows the visual deformation of the Convex Hull structure as the blob shrinks and new city nodes are added to the list. Note that the blob shrinks simultaneously from all directions and the order of insertion is related to both the proximity of the point from the periphery of the blob and the distance between two outer stimuli at the current periphery of the blob where a concavity forms (discussed further in Section \ref{concavity}). The actual order of insertion of cities in this example is given in Fig.~\ref{fig:tour_construct}. 

\begin{figure}[!tbp]
\centering
$$
\begin{array}{cccccccccccccccccccc}
 M &  &  & T &  & D & K &  &  & G &  &  & S & N & Q &  &   &   &  &  \\  
 M &  &  & T &  & D & K &  &  & G &  &  & S & N & Q &  &   &   &  & \textbf{R} \\  
 M & \textbf{I} &  & T &  & D & K &  &  & G &  &  & S & N & Q &  &   &   &  & R \\  
 M & I &  & T &  & D & K &  &  & G & \textbf{J} &  & S & N & Q &  &   &   &  & R \\ 
 M & I & \textbf{H} & T &  & D & K &  &  & G & J &  & S & N & Q &  &   &   &  & R \\ 
 M & I & H & T &  & D & K &  & \textbf{C} & G & J &  & S & N & Q &  &   &   &  & R \\ 
 M & I & H & T &  & D & K &  & C & G & J & \textbf{B} & S & N & Q &  &   &   &  & R \\ 
 M & I & H & T & \textbf{E} & D & K &  & C & G & J & B & S & N & Q &  &   &   &  & R \\ 
 M & I & H & T & E & D & K &  & C & G & J & B & S & N & Q & \textbf{F} &   &   &  & R \\ 
 M & I & H & T & E & D & K &  & C & G & J & B & S & N & Q & F &   &   & \textbf{L} & R \\ 
 M & I & H & T & E & D & K & \textbf{P} & C & G & J & B & S & N & Q & F &   &   & L & R \\ 
 M & I & H & T & E & D & K & P & C & G & J & B & S & N & Q & F &   & \textbf{A} & L & R \\
 M & I & H & T & E & D & K & P & C & G & J & B & S & N & Q & F & \textbf{O} & A & L & R \\
\end{array}
$$
\caption{Gradual construction of tour by city insertion during shrinkage process in dataset 16, as visualised in Fig.~\ref{fig:conv_conc_tsp}. Top row shows initial configuration of blob as convex hull. Each row inserts a new city (bold) into the tour as indicated in Fig.~\ref{fig:conv_conc_tsp}.} 
\label{fig:tour_construct}
\end{figure}

As the blob shrinks, concavities form in the periphery of the blob which move inwards to the centre of the blob shape. The concave deformation is a transformation of the Convex Hull ($\mathbf{CH}$) into a Concave Hull ($\mathbf{OH}$). The Concave Hull, the area occupied by --- or the `shape' of --- a set of points is not as simple to define as its convex hull. It is commonly used in Geographical Information Systems (GIS) as the minimum region (or footprint \cite{galton2006region}) occupied by a set of points, which cannot, in some cases, be represented correctly by the convex hull \cite{duckham2008efficient}. The Concave Hull is related to the structures known as $\alpha$-shapes \cite{edelsbrunner1983shape}. The $\alpha$-shape of a set of points, $P$, is an intersection of the complement of all closed discs of radius $1/\alpha$ that includes no points of $P$. An $\alpha$-shape is a convex hull when $\alpha \rightarrow \infty$. When decreasing $\alpha$, the shapes may shrink, develop holes and become disconnected, collapsing to $P$ when $\alpha \rightarrow 0$. A concave hull is non-convex polygon representing area occupied by $P$ and the concave hull is a connected $\alpha$-shape without holes. In contrast to $\alpha$-shapes, the blob (more specifically, the set of points which it covers) does not become disconnected as it shrinks. As the blob adapts its morphology from Convex Hull to TSP is demonstrates increased concavity with decreased area. Although the shrinkage process is automatically stopped when a TSP tour is formed, the process could indeed continue past the TSP. If shrinkage continues then the blob (now adopting a network shape) will approximate the Steiner minimum tree (SMT), the minimum path between all nodes. As demonstrated in \cite{jones2010influences} the additional Steiner nodes in the SMT may be removed by increasing the attractant projection from the data nodes. The material adapts to the increased attractant concentration by removing the Steiner nodes to approximate the Minimum Spanning Tree (MST).

\subsection{Blob TSP Tour as a Waypoint in the Transition From Convex Hull to Spanning Tree}

The insertion process of adding nodes to the Convex Hull reveals an orderly transition to the TSP which continues after further shrinkage, leading to the following finding.

\begin{finding}
The evolution of the blob shape by morphological adaptation is a transition from $\mathbf{CH}$ to $\mathbf{OH}$ to $\mathbf{TSP}$ to $\mathbf{MST}$ to $\mathbf{SMT}$. 
\end{finding}

We do not explicitly include $\alpha$-shapes in this transition since $\alpha$-shapes can include holes and disconnected structures, which do not form in a defect-free shrinking blob. This transition is based on increasing concavity and decreasing area, and encompasses the a blob TSP tour $\mathbf{bTSP}$ as part of the hierarchy. Note that the blob tour $\mathbf{bTSP}$ is only one instance of the set of possible TSP tours $\mathbf{TSP}$ and is not guaranteed to be the minimal tour. The blob TSP tour is only a transient structure --- a waypoint ---  in the natural shrinkage process (we halt the computation at this point merely because we are interested for the purposes of this report).

\begin{figure}[htbp]
 \centering
 
 \subfigure[Nodes]				 	{\includegraphics[width=0.15\textwidth]{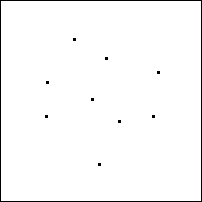}}             
 \subfigure[$\mathbf{DTN}$]	{\includegraphics[width=0.15\textwidth]{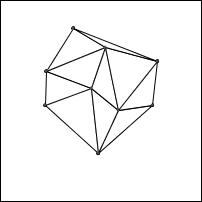}}            
 \subfigure[$\mathbf{GG}$]	{\includegraphics[width=0.15\textwidth]{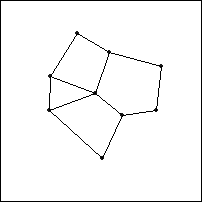}}             
 \subfigure[$\mathbf{RNG}$]	{\includegraphics[width=0.15\textwidth]{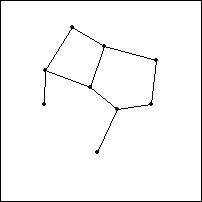}}             
 \subfigure[$\mathbf{MST}$]	{\includegraphics[width=0.15\textwidth]{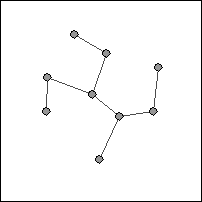}}             
 \subfigure[$\mathbf{SMT}$]	{\includegraphics[width=0.15\textwidth]{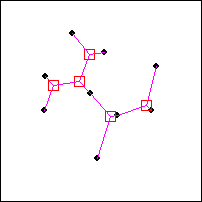}}               
 \subfigure[$\mathbf{CH}$]	{\includegraphics[width=0.15\textwidth]{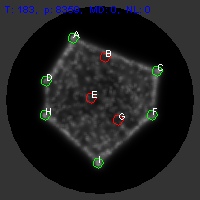}}                
 \subfigure[$\mathbf{OH}$]	{\includegraphics[width=0.15\textwidth]{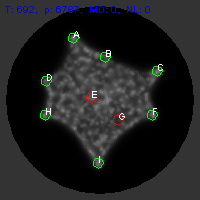}}                
 \subfigure[$\mathbf{OH}$]	{\includegraphics[width=0.15\textwidth]{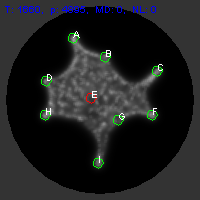}}                
 \subfigure[$\mathbf{bTSP}$]{\includegraphics[width=0.15\textwidth]{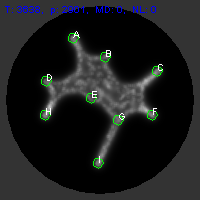}}                
 \subfigure[$\mathbf{MST}$]	{\includegraphics[width=0.15\textwidth]{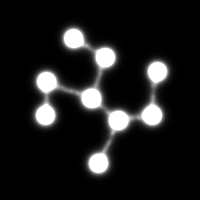}}                
 \subfigure[$\mathbf{SMT}$]	{\includegraphics[width=0.15\textwidth]{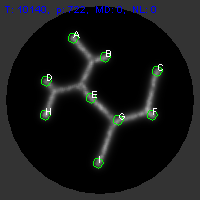}}                  
 \caption[]{Comparison of the Toussaint Hierarchy (top row) and Shrinking Blob Hierarchy (bottom row). (a) Initial source nodes, (b) Delaunay Triangulation (DTN), (c) Gabriel Graph (GG), (d) Relative Neighbourhood Graph (RNG), (e) Minimum Spanning Tree (MST), (f) Steiner Minimum Tree (SMT), (g) Initial blob is patterned as a Convex Hull (CH), (h-i) as the blob shrinks it adopts the Concave Hull, (j) after uncovering the last node a TSP tour is formed, (k) blob can be forced to adopt MST by increasing node concentration, (l) the `natural' end point of blob shrinkage is the SMT.} 
\label{fig:compare_hierarchy}
\end{figure}

It is known from Toussaint that there is a hierarchy of proximity graphs (graphs where edges between points are linked depending on measures of neighbourhood and closeness) \cite{toussaint_1980}. Each member of the hierarchy adds edges and subsumes the edges of lower stages in the hierarchy, and some common graphs (see Fig. \ref{fig:compare_hierarchy}a-e) include the Delaunay triangulation $\mathbf{DTN}$ to Gabriel Graph $\mathbf{GG}$ to Relative Neighbourhood Graph $\mathbf{RNG}$ to Minimum Spanning Tree $\mathbf{MST}$. Also shown is the shortest possible tree between all nodes formed by adding extra Steiner nodes (Fig. \ref{fig:compare_hierarchy}f). It was found in \cite{adamatzky_toussaint} that \emph{Physarum} approximates the Toussaint hierarchy of proximity graphs as it constructs transport networks during its foraging and it was demonstrated in \cite{jones2010influences} that multi-agent transport networks mimicking the behaviour of \emph{Physarum} also minimise these proximity graphs by following this hierarchy in its downwards direction. From a biological perspective traversing the Toussaint hierarchy suggests a mechanism by which \emph{Physarum} can exploit the trade-off between foraging efficiency (many network links) and transport efficiency (fewer but fault tolerant transport links). This mechanism, may also be present in terms of maximising foraging area searched (exploration) and minimising area for efficient transport (exploitation), as suggested in \cite{gunji2011adaptive}. We suggest that the hierarchy we observed in the shrinking blob from $\mathbf{CH}$ to $\mathbf{OH}$ to $\mathbf{TSP}$ to $\mathbf{MST}$ to $\mathbf{SMT}$ may encompass such an area-based exploration-exploitation mechanism (Fig. \ref{fig:compare_hierarchy}g-l). It is notable that there is some overlap between the Toussaint hierarchy and the shrinking blob hierarchy where deepening concavities in the blob hierarchy appear to correspond to the deletion of outer edges in the Toussaint hierarchy, suggesting that there may be some formal relationship between the two. This possible relationship may suggest further studies.

\subsection{Variations in Performance of the Shrinking Blob Method}
\label{concavity}

The results of the shrinking blob method show variations in performance from very good approximations of close-to-minimum tours (Fig.~\ref{fig:results_good}) to less favourable fractions of the minimal tour (Fig.~\ref{fig:results_not_good}). What is the reason for the disparity in performance on these datasets? If we examine the tour paths we can glean some clues as to the difference in performance. In the `good' results examples the major concave regions of the tour formed by the blob closely match the concavities in the exact computed TSP tour (e.g. Fig.~\ref{fig:results_good}a and b). However in the `poor' approximation results we can see that the major concave regions of the blob tour do not match the major concavities in the respective exact computed tours (e.g. Fig.~\ref{fig:results_not_good}a and b). Given that these concave regions are formed from the deformation of the initial Convex Hull we can see that the concavities in the blob tour appear to be formed, and deepened where there are larger distances between the cities on the initial Convex Hull.

\begin{figure}[htbp]
 \centering
 \subfigure[t=100]{\includegraphics[width=0.19\textwidth]{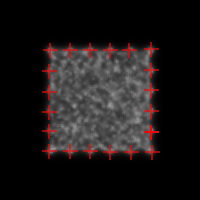}}
 \subfigure[t=3000]{\includegraphics[width=0.19\textwidth]{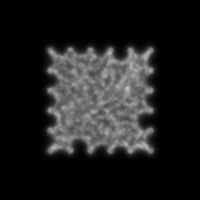}}
 \subfigure[t=3500]{\includegraphics[width=0.19\textwidth]{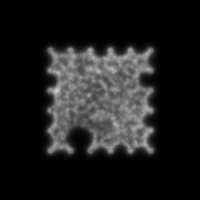}}
 \subfigure[t=4000]{\includegraphics[width=0.19\textwidth]{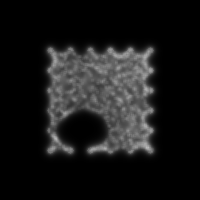}}
 \subfigure[t=5000]{\includegraphics[width=0.19\textwidth]{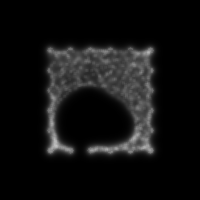}}  
 \subfigure[t=100]{\includegraphics[width=0.19\textwidth]{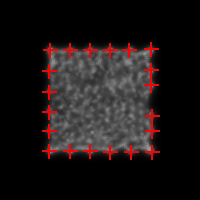}}   
 \subfigure[t=500]{\includegraphics[width=0.19\textwidth]{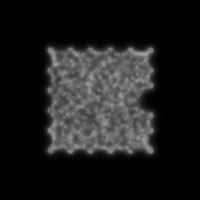}}   
 \subfigure[t=1000]{\includegraphics[width=0.19\textwidth]{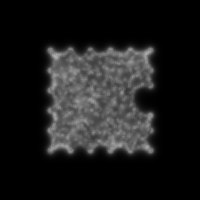}}   
 \subfigure[t=2000]{\includegraphics[width=0.19\textwidth]{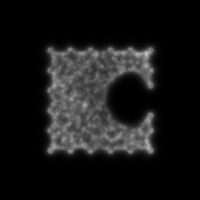}}   
 \subfigure[t=3000]{\includegraphics[width=0.19\textwidth]{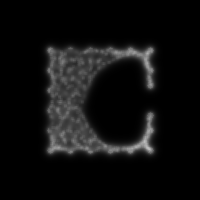}}       
 \subfigure[t=100]{\includegraphics[width=0.19\textwidth]{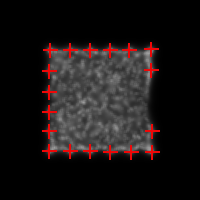}}     
 \subfigure[t=500]{\includegraphics[width=0.19\textwidth]{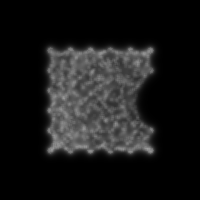}}     
 \subfigure[t=1000]{\includegraphics[width=0.19\textwidth]{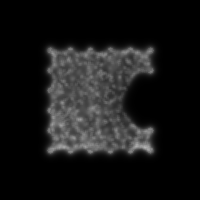}}     
 \subfigure[t=2000]{\includegraphics[width=0.19\textwidth]{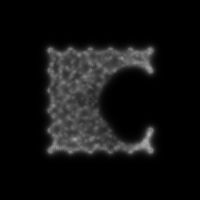}}     
 \subfigure[t=3000]{\includegraphics[width=0.19\textwidth]{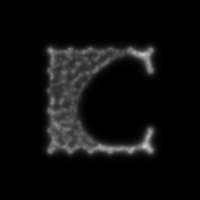}}   
 \subfigure[t=100]{\includegraphics[width=0.19\textwidth]{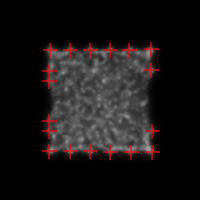}}            
 \subfigure[t=500]{\includegraphics[width=0.19\textwidth]{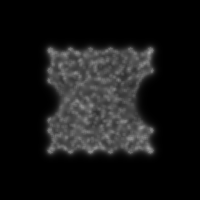}}            
 \subfigure[t=1000]{\includegraphics[width=0.19\textwidth]{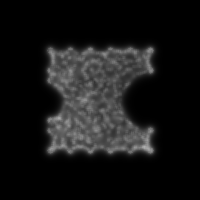}}            
 \subfigure[t=2000]{\includegraphics[width=0.19\textwidth]{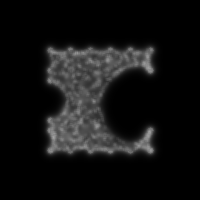}}            
 \subfigure[t=3000]{\includegraphics[width=0.19\textwidth]{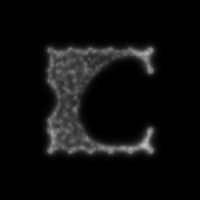}}                
 \caption[]{Distance between nodes affects position and speed of concavity formation in shrinking blob. Nodes arranged in the border of square (indicated by crosses on leftmost images). (a-e) all nodes have identical distance of 20 pixels, (f-j) right side has node gap of 30 pixels, (k-0) right side node gap 60 pixels, (p-t) left side node gap 40 pixels, right side node gap 60 pixels.} 
\label{fig:gap_distance}
\end{figure}

To explore the role of distance on concavity formation we patterned a blob into a square shape by placing regularly placed stimuli around the border of a square (Fig.~\ref{fig:gap_distance}a, stimuli positions, 20 pixels apart, indicated by crosses). When shrinkage of the blob was initiated there is no difference between the stimuli distances. All regions between stimuli initially show small concavities (the `perforations' in Fig.~\ref{fig:gap_distance}b) until one gradually predominates and extends inwards. Also of note is the fact that when one concave region predominates, the other concavities shrink (Fig.~\ref{fig:gap_distance}c-e). The position of the initial dominating concavity is different in each run (presumably due to stochastic influences on the collective material properties of the blob) and this may explain the small differences in performance on separate runs using the same dataset.

When there is a larger gap between stimulus points the predominating concavity forms more quickly and is larger. This is shown in Fig.~\ref{fig:gap_distance}f-j which has a gap of only 30 pixels between neighbouring stimulus points on the right side of the square and in Fig.~\ref{fig:gap_distance}k-o which has a gap of 60 pixels between neighbouring stimulus points. The shorter distance between points in (f-j) generates more tension in the sheet, prevent its deformation. The larger distance between stimuli in (k-o) results in less tension in the sheet at this region and the sheet deforms to generate the concavity.

When there are multiple instances of large distances between stimuli there is competition between the concave regions and the larger region predominates. This is demonstrated in Fig.~\ref{fig:gap_distance}p-t which has a distance gap of 40 pixels on the left side of the square and 60 pixels on the right side. Although two concave regions are formed, the larger deepens whilst the smaller concave region actually shrinks as the blob adapts its shape.

The synthetic examples illustrate the influence of city distance on concavity formation and evolution and these effects are more complex when irregular arrangements of city nodes are used. This is because arrangements of cities present stimuli to the blob sheet when partially uncovered, acting to anchor the blob at these regions, and the morphological adaptation of the blob is thus dynamically affected by the changing spatial configuration of uncovered city nodes. In the examples of relatively poor approximation of the minimum tour (Fig.~\ref{fig:results_not_good}) the initial incorrect selection of concavities are subsequently deepened by the shrinking process, resulting in tours which differ significantly in both their visual shape and in the city order from the optimum tour. In the examples of good comparative performance with the exact solver the blob tours differ only in a small number of nodes.  

\begin{figure}[htbp]
 \centering
 \subfigure[t=44]					{\includegraphics[width=0.19\textwidth]{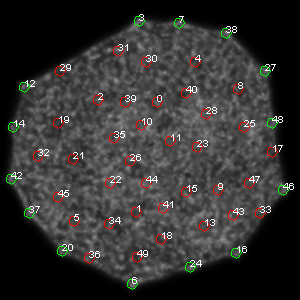}}             
 \subfigure[t=2020]				{\includegraphics[width=0.19\textwidth]{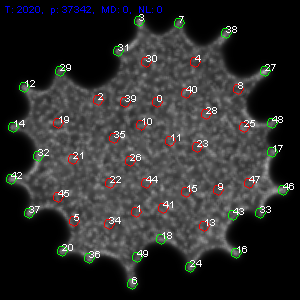}}             
 \subfigure[t=6710]				{\includegraphics[width=0.19\textwidth]{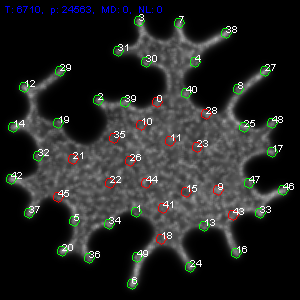}}             
 \subfigure[t=13580]			{\includegraphics[width=0.19\textwidth]{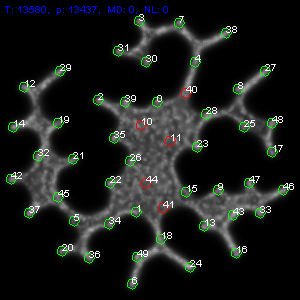}}             
 \subfigure[t=22590]			{\includegraphics[width=0.19\textwidth]{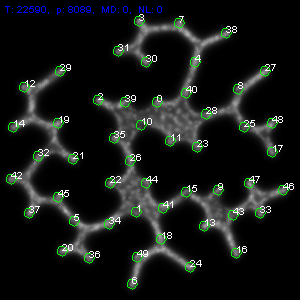}}                 
 \subfigure[Blob 1045]		{\includegraphics[width=0.3\textwidth]{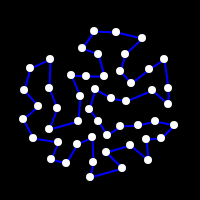}}            
 \subfigure[Concorde 982]	{\includegraphics[width=0.3\textwidth]{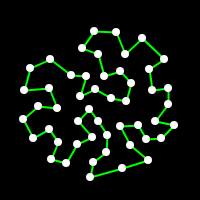}}                
 \caption[]{The performance of the shrinking blob method on a larger dataset. (a-e) evolution of the shrinking blob on 50 node dataset, (f) tour formed by shrinking blob, (g) exact minimum tour by Concorde solver.} 
\label{fig:50_nodes}
\end{figure}

Although outright performance is not the focus of this report, we tested the blob method on a randomly generated dataset of 50 nodes in a preliminary assessment of scalability. We found that over 6 runs the blob method found tours with a best of 1.064, worst of 1.099 and mean of 1.076 compared as a fraction of tour length to the exact minimum computed by the TSP solver, Fig. \ref{fig:50_nodes}. However, these results may also be subject to the variability in performance seen in the 20 node examples.

\section{Discussion}
\label{sec:discussion}

We have presented a simple material-based approach to the computation of the Travelling Salesman Problem using a shrinking blob. The method utilises the emergent morphological adaptation properties of a virtual material arising from local interactions within a multi-agent particle system. We shrink this material over time and its deformation and adaptation to the projected data points yields a tour of the TSP. We should again emphasise that the method is notable for its simplicity and novelty rather than its performance. Indeed the performance, when compared to exact TSP solvers or leading heuristic methods, compares relatively unfavourably in terms of absolute tour distance. The method does, however, contain a number of properties that are intriguing. Firstly, unlike many other nature inspired approaches, the method is not population based and the blob only computes a single instance of a TSP tour. In addition, no attempt is made to modify or optimise the tour. The benefits of population based approaches are that very large search spaces in combinatorial optimisation problems can be traversed and candidate solutions can be compared in some way. This allows the efficient pooling of good solutions, generation of new candidate solutions and avoids local minima. The blob method does not contain any of these beneficial features.

How then, can a shrinking blob naturally generate a good quality (albeit non optimal) TSP tour? The intrinsic performance of the blob is based on its (virtual) material properties which exhibit innate minimisation behaviour. Previous research has demonstrated the network minimisation properties of the material approach, reproducing phenomena seen in soap film evolution \cite{jones2010influences} and lipid nanotube networks \cite{lobovkina2008shape}. The maintenance of uniform shape during blob shrinkage also allows no crossing over of paths and it is known from \cite{flood1956traveling} that crossing paths produce non-optimal tours. More specifically, the tour is constructed by insertion of cities into the list as concave regions are formed in the initial Convex Hull pattern and subsequently deepened. This is similar to algorithmic heuristics which, beginning with a Convex Hull, add cities to the list based on certain cost criteria \cite{golden1980approximate}, \cite{kurz2011heuristics}. In the case of the blob, however, there is no explicit consideration of cost when adding cities to the tour. The mechanism of insertion selection in the blob (by deepening concavities) is intrinsic to its quasi-mechanical properties of the `material' which are influenced by the depth of the city to the Convex Hull boundary and the span distance between boundary stimuli.

Although the blob approach differs from population based nature-inspired heuristics it is reminiscent, in character if not in direct operation, with other analogue based methods. In the Elastic Net algorithm, introduced by Durbin and Willshaw, a circular band is initialised at the approximate centre of a pattern of source TSP nodes. The band is expanded iteratively whilst two forces are applied to points on the band which attempt to minimise distances between cities on the band and the overall length of the band itself \cite{durbin1987analogue}. In the conceptually opposite approach of \cite{al1998efficient} the band is initialised on the Convex Hull and the two forces attempt to constrict the band whilst attracting the band towards a city. The tour formed by the band is then subject to a second `non-deterministic improvement' algorithm to escape local minima. The main difference between the blob method and these `band' approaches is that the material properties of the blob method are an emergent property of, and are distributed within, local interactions between components of the material. The computation is thus an embodied property of the material itself.

Can the mechanism underlying the simple material approximation of the TSP in the blob approach contribute to the question of human performance on the TSP? MacGregor and Ormerod noted that humans produced efficient results on the TSP \cite{macgregor1996human} and this finding stimulated further research into human performance on the problem and possible  perceptual and cognitive mechanisms. In their analysis of experimental findings using human subjects Ormerod and Chronicle noted that global perceptual influences appear to play a role in human approximation of the TSP \cite{ormerod1999global}. MacGregor et al. suggested a model based on insertion of cities into the Convex Hull \cite{macgregor2000model}. This model is similar to the shrinking blob mechanism except that in the blob approach the addition of cities occurs in parallel whereas in the MacGregor et al. model it is a sequential process. The blob method also exhibits another property found in optimal tours, that boundary points in the original convex hull are connected in sequence (the sequence may, of course, be interrupted by interior points). Other competing models to explain human TSP performance exist, including variants of hierarchical pyramid models \cite{graham2000traveling}, \cite{pizlo2006traveling} and the global-local model proposed by Best \cite{best2005model}. Merits, problems, biological plausibility and the role of local vs. global perceptual processes of the competing models have been the subject of lively debate and an assessment is beyond the scope of this paper, but see the review in \cite{macgregor2011human} for an overview. In this review MacGregor states that, despite the growing interest in research into the human performance on the TSP, and combinatorial optimisation problems in general: ``As yet, no algorithms have been put forward to explain performance on the MSTP [Minimum Spanning Tree Problem] and the GSTP [Generalised Steiner Tree Problem]\ldots''. It is notable that the shrinking blob method incorporates approximations of all three problems and executes a natural transition from global to local `perception' using material properties which emerge from very simple low-level and bottom-up interactions. Whether this natural computation employed by the blob is of interest, or utility, to human combinatorial optimisation problems is, however, an open question.

Limitations of the approach, as it stands, include its noted relatively modest performance and the reliance of a manual method to interpret the result of the blob computation. Manual interpretation of the blob tour is, of course, open to experimenter bias and for this reason a methodical process must be followed, as described in section \ref{read_results}. An automated method of tracking the perimeter and `reading' the result of the blob tour would, nevertheless, be of benefit. Further work, including a comprehensive evaluation of model parameters affecting the material properties of the blob may suggest methods by which the basic features of the shrinking blob approach may be adapted, or improved, to improve the performance in comparison with leading heuristic methods. The material properties and computation of the blob emerge from a population of simple multi-agent particles and It would be satisfying if this virtual material could be implemented and embodied in a real physical substrate with the desired physical (for example visco-elastic, free energy minimisation) properties. Alternately it may be possible to translate the material operation of the unconventional computation blob method into a classical algorithmic method.

\section{Acknowledgements} \noindent This authors AA and JJ were supported by the EU research project ``Physarum Chip: Growing Computers from Slime Mould'' (FP7 ICT Ref 316366) and JJ was supported by the SPUR grant award for the project ``Developing non-neural models of material computation in cellular tissues'' from UWE.

\section{Appendix: Shrinking Blob Particle Model Description}
\label{appendix}

The multi-agent particle approach to generate the behaviour of the virtual material blob uses a population of coupled mobile particles with very simple behaviours, residing within a 2D diffusive lattice. Lattice size was $200 \times 200$ pixels. The lattice stores particle positions and the concentration of a local diffusive factor referred to generically as chemoattractant. Collective particle positions represent the global pattern of the blob. The particles act independently and iteration of the particle population is performed randomly to avoid any artifacts from sequential ordering.

\subsection{Generation of Emergent Blob Cohesion and Shape Adaptation}

\begin{figure}[!tbp]
 \begin{center}
  \includegraphics[width=0.8\textwidth]{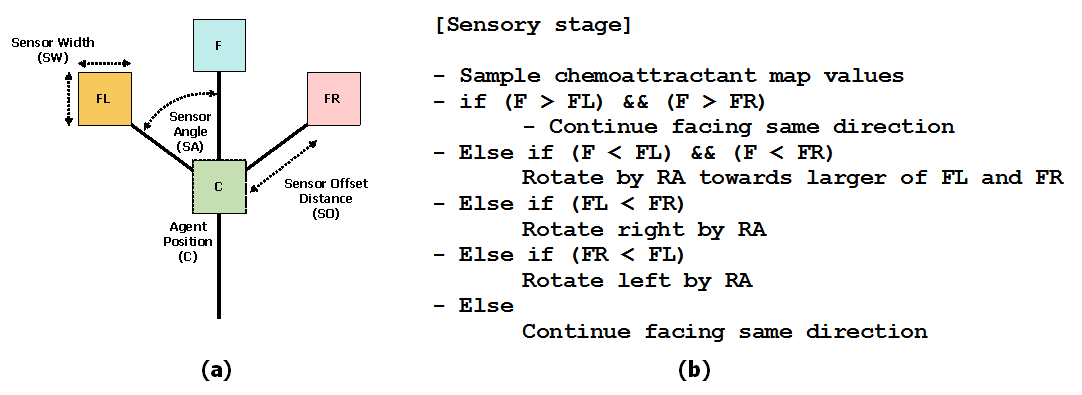}
 \end{center}
 \caption{Architecture of a single component of the shrinking blob and its sensory algorithm. (a) Morphology showing agent position `C' and offset sensor positions (FL, F, FR), (b) Algorithm for particle sensory stage.}
	\label{fig_particle_layout}
\end{figure}

The behaviour of the particles occurs in two distinct stages, the sensory stage and the motor stage. In the sensory stage, the particles sample their local environment using three forward biased sensors whose angle from the forwards position (the sensor angle parameter, SA), and distance (sensor offset, SO) may be parametrically adjusted (Fig.~\ref {fig_particle_layout}a). The offset sensors generate local coupling of sensory inputs and movement to generate the cohesion of the blob. The SO distance is measured in pixels and a minimum distance of 3 pixels is required for strong local coupling to occur. During the sensory stage each particle changes its orientation to rotate (via the parameter rotation angle, RA) towards the strongest local source of chemoattractant (Fig.~\ref{fig_particle_layout}b). After the sensory stage, each particle executes the motor stage and attempts to move forwards in its current orientation (an angle from 0--360 degrees) by a single pixel forwards. Each lattice site may only store a single particle and particles deposit chemoattractant into the lattice only in the event of a successful forwards movement. If the next chosen site is already occupied by another particle move is abandoned and the particle selects a new randomly chosen direction.

\subsection{TSP Problem Representation}

Twenty datasets were generated, each consisting of 20 randomly chosen data points within a circular arena. A condition was added that a minimum distance of 25 pixels must exist between data points. This gives a more uniform distribution of node points, preventing clustering of node points often found in real-world TSP instances, for example those based on real city locations. This was done partly to aid visual tracking of the tour path but also because a less clustered node distribution is known to provide more challenging problem instances (e.g. in human performance on TSP \cite{hirtle1992heuristic}) since there is less likelihood of providing pre-existing cues to intuitive solutions (for example nearest neighbour grouping). Each dataset was saved to a text file. TSP city data points for each dataset were loaded from the text files and were represented by projection of chemoattractant to the diffusion lattice at locations corresponding to a $3 \times 3$ window centred about their $x,y$ position. The projection concentration was 1.275 units per scheduler step. If a node was covered by a portion of the blob (i.e. if the number of agents in a $3 \times 3$ window surrounding the node was $>0$) the projection was reduced to 0.01275 units. Suppression of projection from covered sites was necessary to ensure a uniform concentration within the blob at internal data points. Uncovering of the data points by the shrinking blob acted to increase concentration at exposed nodes, causing the blob to be anchored by the nodes. Diffusion in the lattice was implemented at each scheduler step and at every site in the lattice via a simple mean filter of kernel size $3 \times 3$. Damping of the diffusion distance, which limits the distance of chemoattractant gradient diffusion, was achieved by multiplying the mean kernel value by $0.95$ per scheduler step. 

The blob was initialised by creating a population of particles and inoculating the population within the bounds of a Convex Hull formed by the data points, generate at the start of the experiment by a conventional algorithm. Why choose the Convex Hull as the initial blob pattern? It is possible to select any pattern for the blob shape, for example a circular blob. However, selecting any arbitrary pattern to cover the points might leave a large portion of the blob overlapping vacant space. This could result in shrinkage in one part of the blob taking place over vacant space with shrinkage in other parts over city points and would bias the output result. The Convex Hull was thus chosen as a simple representation of the perimeter of the data points. The exact population size differed depending on the area of the Convex Hull but the initial blob was typically composed of between 10000 and 15000 particles. Particles were given random initial positions within these confines and random initial orientations. Any particles migrating out of the bounds of the Convex Hull region were deleted. Particle sensor offset (SO) was 7 pixels. Angle of rotation (RA) and sensor angle (SA) were both set to 60 degrees in all experiments. Agent forward displacement was 1 pixel per scheduler step and particles moving forwards successfully deposited 5 units of chemoattractant into the diffusion lattice. Both data projection stimuli and agent particle trails were represented by the same chemoattractant ensuring that the particles were attracted to both data stimuli and other agents' trails. The collective behaviour of the particle population was cohesion and morphological adaptation to the configuration of stimuli.

\subsection{Shrinkage Mechanism}
 
Adaptation of the blob size is implemented via tests at regular intervals as follows. If there are 1 to 10 particles in a $9 \times 9$ neighbourhood of a particle, and the particle has moved forwards successfully, the particle attempts to divide into two if there is a space available at a randomly selected empty location in the immediate $3 \times 3$ neighbourhood surrounding the particle. If there are 0 to 80 particles in a $9 \times 9$ neighbourhood of a particle the particle survives, otherwise it is deleted. Deletion of a particle leaves a vacant space at this location which is filled by nearby particles, causing the blob to shrink slightly. As the process continues the blob shrinks and adapts to the stimuli provided by the configuration of city data points. The frequency at which the growth and shrinkage of the population is executed determines a turnover rate for the particles. The frequency of testing for particle division was every 5 scheduler steps and the frequency for testing for particle removal was every 10 scheduler steps. Since the shrinking blob method is only concerned with the reduction in size of the population it might be asked as to why there are tests for particle division at all. The particle division mechanism is present to ensure that the adaptation of the blob sheet is uniform across the sheet to prevent `tears' or holes forming within the blob sheet, particularly at the start of an experiment when flux within the blob is initially established.

\subsection{Halting Mechanism}

The shrinkage of the blob was halted when all data points were partially uncovered by using the following calculation. At each data point the number of particles surrounding the point in a $5 \times 5$ window was sampled. If the number of particles in this window was $< 15$ the node was classified as partially uncovered. When all nodes are uncovered the model is halted and a greyscale representation of the diffusive lattice is saved to disk to read the result of the blob tour. The calculated tour path was saved in a text file and tour distance was calculated. The exact minimum tour path for each dataset was calculated for comparison by loading the data point configuration files into the Concorde TSP solver \cite{applegate2006concorde}.

\bibliography{references}

\begin{thebibliography}{10}

\bibitem{adamatzky2007kum}
A.~Adamatzky.
\newblock \emph{Physarum Machine}: Implementation of a {Kolmogorov-Uspensky}
  machine on a biological substrate.
\newblock {\em Parallel Processing Letters}, 17(4):455--467, 2007.

\bibitem{adamatzky_toussaint}
A.~Adamatzky.
\newblock Developing proximity graphs by \emph{Physarum polycephalum}: does the
  plasmodium follow the toussaint hierarchy.
\newblock {\em Parallel Processing Letters}, 19:105--127, 2008.

\bibitem{adamatzky_physarumgate}
A.~Adamatzky.
\newblock Slime mould logical gates: exploring ballistic approach.
\newblock {\em Arxiv preprint arXiv:1005.2301}, 2010.

\bibitem{adamatzky2011planarshape}
A.~Adamatzky.
\newblock Slime mould computes planar shapes.
\newblock {\em Arxiv preprint arXiv:1106.0305}, 2011.

\bibitem{adamatzky_jones_NC}
A.~Adamatzky and J.~Jones.
\newblock Programmable reconfiguration of \emph{Physarum} machines.
\newblock {\em Natural Computing}, 9(1):219--237, 2010.

\bibitem{al1998efficient}
M.~Al-Mulhem and T.~Al-Maghrabi.
\newblock Efficient convex-elastic net algorithm to solve the euclidean
  traveling salesman problem.
\newblock {\em Systems, Man, and Cybernetics, Part B: Cybernetics, IEEE
  Transactions on}, 28(4):618--620, 1998.

\bibitem{AonoM07PhysarumNeuroComp}
M.~Aono and M.~Hara.
\newblock Amoeba-based nonequilibrium neurocomputer utilizing fluctuations and
  instability.
\newblock In {\em 6th Int. Conf., UC 2007}, volume 4618 of {\em LNCS}, pages
  41--54, Kingston, Canada, August 13-17 2007. Springer.

\bibitem{aono2008spontaneous}
M.~Aono and M.~Hara.
\newblock Spontaneous deadlock breaking on amoeba-based neurocomputer.
\newblock {\em BioSystems}, 91(1):83--93, 2008.

\bibitem{aono_neurophys}
M.~Aono, Y.~Hirata, M.~Hara, and K.~Aihara.
\newblock Amoeba-based chaotic neurocomputing: Combinatorial optimization by
  coupled biological oscillators.
\newblock {\em New Generation Computing}, 27(2):129--157, 2009.

\bibitem{applegate2006concorde}
D.~Applegate, R.~Bixby, V.~Chvatal, and W.~Cook.
\newblock Concorde tsp solver.
\newblock {\em URL http://www. tsp. gatech. edu/concorde}, 2006.

\bibitem{best2005model}
B.J. Best.
\newblock A model of fast human performance on a computationally hard problem.
\newblock In {\em Proceedings of the 27th annual conference of the cognitive
  science society}, pages 256--262, 2005.

\bibitem{de2004formation}
B.~de~Lacy~Costello, N.~Ratcliffe, A.~Adamatzky, A.L. Zanin, A.W. Liehr, and
  H.G. Purwins.
\newblock The formation of voronoi diagrams in chemical and physical systems:
  experimental findings and theoretical models.
\newblock {\em International journal of bifurcation and chaos in applied
  sciences and engineering}, 14(7):2187--2210, 2004.

\bibitem{dorigoa2000ant}
M.~Dorigo, E.~Bonabeau, and G.~Theraulaz.
\newblock Ant algorithms and stigmergy.
\newblock {\em Future Generation Computer Systems}, 16(8):851--871, 2000.

\bibitem{dry2006human}
Matthew Dry, Michael~D Lee, Douglas Vickers, and Peter Hughes.
\newblock Human performance on visually presented traveling salesperson
  problems with varying numbers of nodes.
\newblock {\em The Journal of Problem Solving}, 1(1):4, 2006.

\bibitem{duckham2008efficient}
M.~Duckham, L.~Kulik, M.~Worboys, and A.~Galton.
\newblock Efficient generation of simple polygons for characterizing the shape
  of a set of points in the plane.
\newblock {\em Pattern Recognition}, 41(10):3224--3236, 2008.

\bibitem{durbin1987analogue}
R.~Durbin and D.~Willshaw.
\newblock An analogue approach to the travelling salesman problem using an
  elastic net method.
\newblock {\em Nature}, 326(6114):689--691, 1987.

\bibitem{edelsbrunner1983shape}
H.~Edelsbrunner, D.~Kirkpatrick, and R.~Seidel.
\newblock On the shape of a set of points in the plane.
\newblock {\em Information Theory, IEEE Transactions on}, 29(4):551--559, 1983.

\bibitem{flood1956traveling}
M.M Flood.
\newblock The traveling-salesman problem.
\newblock {\em Operations Research}, 4(1):61--75, 1956.

\bibitem{galton2006region}
A.~Galton and M.~Duckham.
\newblock What is the region occupied by a set of points?
\newblock {\em Geographic Information Science}, pages 81--98, 2006.

\bibitem{golden1980approximate}
B.~Golden, L.~Bodin, T.~Doyle, and W.~Stewart.
\newblock Approximate traveling salesman algorithms.
\newblock {\em Operations research}, 28(3-Part-II):694--711, 1980.

\bibitem{graham2000traveling}
Scott~M Graham, Anupam Joshi, and Zygmunt Pizlo.
\newblock The traveling salesman problem: A hierarchical model.
\newblock {\em Memory \& Cognition}, 28(7):1191--1204, 2000.

\bibitem{gunji2011adaptive}
Y.-P. Gunji, T.~Shirakawa, T.~Niizato, M.~Yamachiyo, and I.~Tani.
\newblock An adaptive and robust biological network based on the
  vacant-particle transportation model.
\newblock {\em Journal of Theoretical Biology}, 272(1):187--200, 2011.

\bibitem{hasegawa2011verification}
M~Hasegawa.
\newblock Verification and rectification of the physical analogy of simulated
  annealing for the solution of the traveling salesman problem.
\newblock {\em Physical review E}, 83(3):036708, 2011.

\bibitem{hirtle1992heuristic}
S.C. Hirtle and T.~G{\"a}rling.
\newblock Heuristic rules for sequential spatial decisions.
\newblock {\em Geoforum}, 23(2):227--238, 1992.

\bibitem{hopfield1986computing}
J.J. Hopfield and D.W. Tank.
\newblock Computing with neural circuits: A model.
\newblock {\em Science}, 233(4764):625, 1986.

\bibitem{jones_alife_2010}
J.~Jones.
\newblock Characteristics of pattern formation and evolution in approximations
  of \emph{Physarum} transport networks.
\newblock {\em Artificial Life}, 16(2):127--153, 2010.

\bibitem{jones2010influences}
J.~Jones.
\newblock Influences on the formation and evolution of \emph{Physarum
  polycephalum} inspired emergent transport networks.
\newblock {\em Natural Computing}, pages 1--25, 2010.

\bibitem{jones2010emergence}
J.~Jones.
\newblock {The emergence and dynamical evolution of complex transport networks
  from simple low-level behaviours}.
\newblock {\em International Journal of Unconventional Computing},
  6(2):125--144, 2010.

\bibitem{Jones2011reconfig}
J.~Jones.
\newblock Towards programmable smart materials: Dynamical reconfiguration of
  emergent transport networks.
\newblock {\em Int. Journal of Unconventional Comput.}, 7(6):423--447, 2011.

\bibitem{jones2010towards}
J.~Jones and A.~Adamatzky.
\newblock Towards \emph{Physarum} binary adders.
\newblock {\em Biosystems}, 101(1):51--58, 2010.

\bibitem{kurz2011heuristics}
M.E. Kurz.
\newblock Heuristics for the traveling salesman problem.
\newblock {\em Wiley Encyclopedia of Operations Research and Management
  Science}, 2011.

\bibitem{larranaga1999genetic}
P.~Larranaga, C.M.H. Kuijpers, R.H. Murga, I.~Inza, and S.~Dizdarevic.
\newblock Genetic algorithms for the travelling salesman problem: A review of
  representations and operators.
\newblock {\em Artificial Intelligence Review}, 13(2):129--170, 1999.

\bibitem{lihoreau2010travel}
M.~Lihoreau, L.~Chittka, and N.E. Raine.
\newblock Travel optimization by foraging bumblebees through readjustments of
  traplines after discovery of new feeding locations.
\newblock {\em The American Naturalist}, 176(6):744--757, 2010.

\bibitem{lobovkina2008shape}
T.~Lobovkina, P.G. Dommersnes, S.~Tiourine, J.F. Joanny, and O.~Orwar.
\newblock {Shape optimization in lipid nanotube networks}.
\newblock {\em The European Physical Journal E: Soft Matter and Biological
  Physics}, 26(3):295--300, 2008.

\bibitem{macgregor2011human}
J.N. MacGregor and Y.~Chu.
\newblock Human performance on the traveling salesman and related problems: A
  review.
\newblock {\em The Journal of Problem Solving}, 3(2):2, 2011.

\bibitem{macgregor1996human}
J.N. MacGregor and T.~Ormerod.
\newblock Human performance on the traveling salesman problem.
\newblock {\em Attention, Perception, \& Psychophysics}, 58(4):527--539, 1996.

\bibitem{macgregor2000model}
J.N. MacGregor, T.C. Ormerod, and E.P. Chronicle.
\newblock A model of human performance on the traveling salesperson problem.
\newblock {\em Memory \& Cognition}, 28(7):1183--1190, 2000.

\bibitem{nakagaki2007intelligent}
T.~Nakagaki and R.D. Guy.
\newblock Intelligent behaviors of amoeboid movement based on complex dynamics
  of soft matter.
\newblock {\em Soft Matter}, 4(1):57--67, 2007.

\bibitem{NakagakiT04MultFoodSrc}
T.~Nakagaki, R.~Kobayashi, Y.~Nishiura, and T.~Ueda.
\newblock Obtaining multiple separate food sources: behavioural intelligence in
  the \textsl{{P}hysarum} plasmodium.
\newblock {\em R. Soc. Proc.: Biol. Sci.}, 271(1554):2305--2310, 2004.

\bibitem{nakagaki2007effects}
T.~Nakagaki, T.~Saigusa, A.~Tero, and R.~Kobayashi.
\newblock Effects of amount of food on path selection in the transport network
  of an amoeboid organism.
\newblock In {\em Proceedings of the International Symposium on Topological
  Aspects of Critical Systems and Networks. World Scientific}, 2007.

\bibitem{NakagakiT00MazeSolve}
T.~Nakagaki, H.~Yamada, and A.~Toth.
\newblock Intelligence: Maze-solving by an amoeboid organism.
\newblock {\em Nature}, 407:470, 2000.

\bibitem{ormerod1999global}
T.C. Ormerod and E.P. Chronicle.
\newblock Global perceptual processing in problem solving: The case of the
  traveling salesperson.
\newblock {\em Attention, Perception, \& Psychophysics}, 61(6):1227--1238,
  1999.

\bibitem{pizlo2006traveling}
Z.~Pizlo, E.~Stefanov, J.~Saalweachter, Z.~Li, Y.~Haxhimusa, and W.G Kropatsch.
\newblock Traveling salesman problem: A foveating pyramid model.
\newblock {\em The Journal of Problem Solving}, 1(1):8, 2006.

\bibitem{reyes2002glow}
D.R. Reyes, M.M. Ghanem, G.M. Whitesides, and A.~Manz.
\newblock Glow discharge in microfluidic chips for visible analog computing.
\newblock {\em Lab Chip}, 2(2):113--116, 2002.

\bibitem{shirakawa2009planedivision}
T.~Shirakawa, A.~Adamatzky, Y.-P. Gunji, and Y.~Miyake.
\newblock On simultaneous construction of voronoi diagram and delaunay
  triangulation by \emph{Physarum polycephalum}.
\newblock {\em International Journal of Bifurcation and Chaos},
  19(9):3109--3117, 2009.

\bibitem{stepney2008neglected}
S.~Stepney.
\newblock The neglected pillar of material computation.
\newblock {\em Physica D: Nonlinear Phenomena}, 237(9):1157--1164, 2008.

\bibitem{toussaint_1980}
G.T. Toussaint.
\newblock The relative neighbourhood graph of a finite planar set.
\newblock {\em Pattern Recognition}, 12(4):261--268, 1980.

\bibitem{TsudaS04PhysarumComp}
S.~Tsuda, M.~Aono, and Y.-P. Gunji.
\newblock Robust and emergent \textsl{{P}hysarum} logical-computing.
\newblock {\em BioSystems}, 73:45--55, 2004.

\bibitem{zauner1996parallel}
K.P. Zauner and M.~Conrad.
\newblock Parallel computing with dna: toward the anti-universal machine.
\newblock {\em Parallel Problem Solving from Nature PPSN IV}, pages 696--705,
  1996.

\end{thebibliography}
\bibliographystyle{plain}

\end{document}